\documentclass[11pt]{article}

\usepackage{amssymb,float,graphicx}
\usepackage{amsmath,amsfonts} 
\usepackage{bm}
\usepackage{epstopdf}
\usepackage{authblk}
\usepackage{float}

\usepackage{color}

\usepackage{makecell}

\usepackage{hyperref}
\usepackage[margin = 1in]{geometry}

\newcommand{\hide}[1]{}

\newcommand{\ie}{\emph{i.e.}}

\renewcommand{\hat}{\widehat} 
\newcommand{\ubar}{\overline{u}}
\newcommand{\rhobar}{\bar{\rho}}
\newcommand{\sgn}{\mathrm{sgn}}

\newcommand*{\p}[1]{{\partial_{\theta_{#1}}}}

\newcommand{\OO}{\mathcal{O}}
\newcommand{\cO}{\widetilde{\Omega}}
\newcommand{\cOh}{\Omega}

\newcommand{\R}{\mathbb{R}}
\newcommand{\rmd}{\,\mathrm{d}}
\newcommand{\dd}{D}

\newcommand*{\cU}{\mathcal{U}}
\newcommand*{\M}{\mathcal{M}}
\newcommand*{\N}{\mathcal{N}}
\newcommand*{\PP}{\mathcal{P}}
\newcommand*{\FF}{\mathcal{G}^{(\pm)}}
\newcommand*{\D}{\mathcal{D}}
\newcommand*{\QQ}{\mathcal{Q}}

\newcommand{\CPMI}{\Delta_{\rm CPMI}}
\newcommand{\SPMI}{\Delta_{\rm SPMI}}

\usepackage{xcolor}

\title{Whitham modulation theory and two-phase instabilities for
  generalized nonlinear Schr\"{o}dinger equations with full
  dispersion}

\author[1]{Patrick Sprenger}
\author[2]{Mark A. Hoefer}
\author[1]{Boaz Ilan}

\affil[1]{Department of Applied Mathematics, University of California Merced}
\affil[2]{Department of Applied Mathematics, University of Colorado Boulder}

\begin{document}

\maketitle

\begin{abstract}
  The generalized nonlinear Schr\"odinger equation with full
  dispersion (FDNLS) is considered in the semiclassical regime.  The
  Whitham modulation equations are obtained for the FDNLS equation
  with general linear dispersion and a generalized, local
  nonlinearity.  Assuming the existence of a four-parameter family of
  two-phase solutions, a multiple-scales approach yields a system of
  four independent, first order, quasi-linear conservation laws of
  hydrodynamic type that correspond to the slow evolution of the two
  wavenumbers, mass, and momentum of modulated periodic traveling
  waves.  The modulation equations are further analyzed in the
  dispersionless and weakly nonlinear regimes.  The ill-posedness of
  the dispersionless equations corresponds to the classical criterion
  for modulational instability (MI).  For modulations of linear waves,
  ill-posedness coincides with the generalized MI criterion, recently
  identified by Amiranashvili and Tobisch
  \cite{amiranashvili_extended_2019}.  A new instability index is
  identified by the transition from real to complex characteristics
  for the weakly nonlinear modulation equations.  This instability is
  associated with long-wavelength modulations of nonlinear two-phase
  wavetrains and can exist even when the corresponding one-phase
  wavetrain is stable according to the generalized MI criterion.
  Another interpretation is that, while infinitesimal perturbations of
  a periodic wave may not grow, small but finite amplitude
  perturbations may grow, hence this index identifies a nonlinear
  instability mechanism for one-phase waves.  Classifications of
  instability indices for multiple FDNLS equations with higher order dispersion, including applications to finite depth water waves and the discrete NLS equation are presented and compared with direct numerical simulations.
\end{abstract}

%

\section{Introduction}
We study the full-dispersion nonlinear Schr\"odinger (FDNLS) equation
\begin{equation}
\label{eq:generalized-NLS}
 i\psi_t   =   \cO(-i\partial_x)\psi + f'(|\psi|^2)\psi~,
\end{equation}
where the pseudo-differential operator $\cO(-i\partial_x)$ captures full
linear dispersion, whose action is defined through the Fourier
transform as
$$
\cO \left ( -i\partial_x \right ) \psi(x,t) \equiv \frac{1}{2\pi} \int_{\R}
\cOh(\xi) e^{i\xi x}  \int_{\R} \psi(x',t) e^{-i\xi x'} \rmd x' \rmd \xi ~ ,
\quad 
$$
where $\cOh(\xi)$ is a smooth, real-valued dispersion relation.  The
function $f$ in~\eqref{eq:generalized-NLS} is a smooth, generalized
nonlinearity. The classical, cubic NLS equation with second-order dispersion is the
special case of \eqref{eq:generalized-NLS} with
$\cOh(\xi) = \frac{1}{2}\xi^2$ and $f(\rho) = \sigma \rho^2/2$, \ie,
\begin{equation}
  \label{eq:NLS-Classical}
  i\psi_t = - \frac{1}{2} \psi_{xx} + \sigma|\psi|^2\psi , \quad
  \sigma = \pm 1~.
\end{equation}
The NLS equation \eqref{eq:NLS-Classical} is a universal model of
the slowly varying (weakly dispersive) and weakly nonlinear envelope of a monochromatic wave train \cite{ablowitz_nonlinear_2011,newell_solitons_1985}. 

The formulation of the FDNLS equation is similar to the modeling
approach that was pioneered by Whitham in the context of shallow water
waves \cite{whitham_variational_1967}.  The Whitham equation
\begin{equation}
  \label{eq:1}
  u_t + uu_x + \widetilde{\Omega} \left ( -i\partial_x \right ) u = 0,
\end{equation}
incorporates weak nonlinearity and the full linear dispersion of
unidirectional water waves with phase speed
$\Omega(k) = \sqrt{\tanh k}$ and corresponding pseudo-differential
operator defined by the symbol
$\widetilde{\Omega}(-i\partial_x) = \sqrt{\tan(-i\partial_x)}$.
Equation \eqref{eq:1} has been shown to be a superior model when
approximating the Euler equations \cite{moldabayev_whitham_2015} and
experiments \cite{trillo_observation_2016,carter_bidirectional_2018}
compared to the Korteweg-de Vries (KdV) equation in the shallow water
regime where
$\widetilde{\Omega} \left ( -i\partial_x \right ) u \sim u_x +
u_{xxx}/6$. Whitham's idea is natural from a modeling standpoint and
has since been generalized to other physical scenarios and model
equations, primarily in the context of single- or multi-layer fluids
\cite{joseph_solitary_1977,lannes_water_2013,aceves-sanchez_numerical_2013,kharif_nonlinear_2018,dinvay_whitham_2019,hur_modulational_2019,binswanger_whitham_2021}. In
the same spirit, we view the FDNLS equation \eqref{eq:generalized-NLS}
as a generic weakly nonlinear, strongly dispersive modulation equation
for one-phase wavetrains, which has been used in various applications
such as water waves \cite{trulsen2000weakly} and optics
\cite{amiranashvili_extended_2019}.

The NLS equation~\eqref{eq:NLS-Classical} is the canonical model for
weakly-dispersive deep water
waves~\cite{zakharov_stability_1972,ablowitz_nonlinear_2011}. However,
higher-order dispersive effects can be significant for deep water
waves~\cite{dysthe1979note,tulin1999laboratory,trulsen2000weakly,sedletsky2003fourth,slunyaev2005high,lannes_water_2013},
such as for oceanic rogue waves (also known as freak waves or peaking
waves) that exhibit steep, cusp-like
profiles~\cite{dysthe2008oceanic,ankiewicz2014extended}. Higher-order
generalied NLS models are also significant in other applications.  In
particular, higher-order dispersive effects are significant for
ultrashort optical pulses~\cite{agrawal2000nonlinear} where the
propagation of intense laser pulses in optical fibers have been
studied for more than 30 years
(cf.~\cite{potasek1987modulation,agrawal2000nonlinear}).  In that
context, the measured spectrum of optical fibers corresponds to
$\Omega(\xi)$\footnote{For optical pulses, $t$ is propagation
  distance, $x$ is time in the frame moving with pulse's center, $\xi$
  is temporal frequency, and $\Omega(\xi)$ is the carrier
  wavenumber.}.
Example experimental
studies on short-wave effects that are not captured by the NLS
equation \eqref{eq:NLS-Classical} include radiative effects in optical
fibers~\cite{conforti2013dispersive,conforti2014radiative,malaguti_dispersive_2014} and longitudinal soliton tunneling in dispersion-shifted
fibers~\cite{marest_longitudinal_2017}.  Stronger dispersion has been
used to predict the appearance of new types of  optical instabilities
\cite{zhang2010general,amiranashvili_extended_2019}.  

A powerful mathematical tool for analyzing modulations of nonlinear
waves is Whitham modulation theory \cite{whitham_linear_1974}.
Whitham modulation theory describes the slow evolution of nonlinear
wavetrains by a system of quasi-linear, first order partial
differential equations for the wave parameters known as the Whitham
modulation equations---plural so as not to be confused with the
singular Whitham equation \eqref{eq:1}.  The Whitham equations are
used to describe long wavelength modulations of strongly nonlinear
wavetrains, hence can be viewed as a large amplitude generalization of
the classical NLS equation \eqref{eq:NLS-Classical}, albeit without
dispersive corrections \cite{newell_solitons_1985}.
One of the earliest applications of modulation theory was to the studies of modulational instability (MI) of nonlinear periodic traveling waves  in optics \cite{talanov1965self,bespalov_filamentary_1966,ostrovskii1967propagation} and water waves
\cite{benjamin_disintegration_1967,benjamin_instability_1967}.
Around the same time, it was recognized that the ellipticity, hence
ill-posedness of the initial value problem, of the Whitham modulation
equations implies MI \cite{whitham_linear_1974}.  See
\cite{zakharov_modulation_2009} for a more detailed history.

In this work, we consider long wavelength modulations of both one- and
two-phase solutions of the FDNLS equation \eqref{eq:generalized-NLS}.
Assuming the existence of two-phase solutions, we obtain the Whitham
modulation equations in a general form for slow spatio-temporal
variation of the solution's parameters.  We then obtain approximate
two-phase solutions and analyze the corresponding modulation equations
in more detail.  We determine their hyperbolicity and a new
\textit{two-phase modulational instability index} that determines
whether small amplitude, long-wavelength perturbations of the
two-phase solution grow.  When the dispersion $\cOh(\xi)$ is cubic,
quartic, from finite depth water waves, or from the discrete NLS
equation, we identify regimes where the periodic, one-phase solution
is modulationally stable but the two-phase solution is modulationally
unstable.  This result implies nonlinear instability, i.e., that
appropriately selected small but finite amplitude perturbations of the
one-phase solution grow even though infinitesimal, linear
perturbations do not.

In order to set the stage for the modulation and stability analysis of
FDNLS solutions, we now briefly review modulation theory and
modulational instability for the cubic NLS equation
\eqref{eq:NLS-Classical}.

\subsection{Modulation theory}
\label{sec:one-phase-modul}

The simplest one-phase solution of
Eq.~\eqref{eq:NLS-Classical} is
\begin{align}\label{eq:stokes_NLS}
  \psi(x,t) = \sqrt{\rhobar} e^{i \theta}
  , \quad \theta = \ubar x - \gamma t, \quad \gamma =
  \frac{1}{2}\ubar^2 + \sigma \rhobar, \quad 
  \rhobar > 0, \quad \ubar \in \mathbb{R},
\end{align}
where $\rhobar = |\psi|^2$ is the square modulus,
$\ubar = i (\psi_x^* \psi - \psi_x\psi^*)/\rhobar$ is the wavenumber
($\psi^*$ is the complex conjugate of $\psi$), and $\gamma$ is the
frequency.  These quantities admit a hydrodynamic interpretation in
which $\rhobar$ and $\ubar$ are analogous to the fluid density and
velocity, respectively \cite{jin_semiclassical_1999}.  The solution
\eqref{eq:stokes_NLS} is often referred to as the Stokes wave, owing
to its derivation by Stokes in the context of weakly nonlinear water
waves \cite{stokes_theory_1847}.

Modulation equations for the Stokes solution's parameters
$(\rhobar,\ubar)$ can be obtained with the WKB-like, multiscale ansatz
\cite{whitham_linear_1974}
\begin{equation}
  \label{eq:4}
  \psi(x,t) = \sqrt{\rhobar(X,T)} e^{i\theta} + \epsilon
  \psi_1(\theta,X,T) + \cdots, 
\end{equation}
where $X = \epsilon x$, $T = \epsilon t$ and
\begin{equation}
  \label{eq:5}
  \theta = S/\epsilon, \quad \theta_x = S_X = \ubar(X,T) , \quad
  \theta_t = S_T =  -\gamma(X,T), \quad 0 < \epsilon \ll 1.
\end{equation}
This ansatz is then inserted into Eq.~\eqref{eq:NLS-Classical} and
like powers of $\epsilon$ are equated.  At leading order, we obtain the
same frequency relation as in \eqref{eq:stokes_NLS}, but now
applicable locally
$\gamma(X,T) = \tfrac12 \ubar(X,T)^2 + \sigma \rhobar(X,T)$.
Combining solvability at $\mathcal{O}(\epsilon)$ over the space of
$2\pi$-periodic functions in $\theta$ for $\psi_1(\theta,X,T)$ and the
compatibility condition $S_{XT} = S_{TX}$ yields the
modulation equations, also known as the \emph{shallow water equations},
\begin{subequations}
  \label{eq:disperionless_modeq}
  \begin{align}
    \rhobar_T + (\rhobar \ubar)_X &= 0, \\[2mm]
    \ubar_T + \left (\tfrac12  \ubar^2 + \sigma \rhobar \right )_X &= 0~.
  \end{align}
\end{subequations}
The characteristic velocities of the shallow water equations
\eqref{eq:disperionless_modeq} are
\begin{align}
  \label{eq:disperionless_velocities}
  \lambda_{1,2} &= \ubar \pm \sqrt{\sigma \rhobar}~.    
\end{align}
Equations~\eqref{eq:disperionless_modeq} describe large amplitude,
nonlinear modulations.  When $\sigma = 1$, the modulation equations
are hyperbolic.  As such, smooth initial data can develop a gradient
catastrophe in finite time, which is regularized by higher-order
dispersive effects not included in \eqref{eq:disperionless_modeq}.
This regularization can be achieved by incorporating an additional
phase into modulation theory that results in the formation of a
dispersive shock wave \cite{el_decay_1995,el_general_1995}.  When
$\sigma = -1$, the characteristic
velocities~\eqref{eq:disperionless_velocities} are complex and the
initial value problem is ill-posed. Nevertheless, the evolution of
certain initial data that leads to singularity formation can also be
regularized by appealing to additional modulation phases
\cite{el_modulational_1993,el_dam_2016-1,biondini_riemann_2018}.

The NLS equation \eqref{eq:NLS-Classical} admits exact solutions in
terms of Jacobi elliptic functions. The solutions take the form
\begin{equation}
  \label{eq:3}
  \psi(x,t) = \sqrt{\rho(\theta_2)}\exp\left(i \theta_1 + i
    \int_0^{\theta_2} u(s) ds\right),
\end{equation}
where $\theta_1 = \ubar x - \gamma t$, $\theta_2 = kx - \omega t$,
$\rho(\theta)$ is a $2\pi$-periodic function, and $u(\theta)$ is a
mean-zero, $2\pi$-periodic function.  The form of the solutions for
$\sigma = \pm 1$ are well documented, see, e.g.,
\cite{kamchatnov_nonlinear_2000}. Since \eqref{eq:3} is periodic in
two independent phases $\theta_1$ and $\theta_2$, it is called a
\emph{two-phase solution}. The phase $\theta_1$ is sometimes referred
to as trivial because it corresponds to the Stokes wave background
\eqref{eq:stokes_NLS} when $\rho(\theta) = \rhobar$ and
$u(\theta) = 0$ are constant.

The modulation equations for the parameters of the two-phase solution family \eqref{eq:3} have been derived in both cases $\sigma = \pm 1$
\cite{forest1986geometry,pavlov_nonlinear_1987}.  They can be cast in the diagonalized form
\begin{equation}
  \label{eq:6}
  \frac{\partial r_j}{\partial T} + \lambda_j(\mathbf{r})
  \frac{\partial r_j}{\partial X} = 0, \quad j = 1,2,3,4 ,
\end{equation}
where $\mathbf{r} = (r_1,r_2,r_3,r_4)$ is the vector of parameters for the two-phase solution \eqref{eq:3}.  When $\sigma = -1$, the characteristic velocities $\lambda_j$ are generically complex, hence \eqref{eq:6} is elliptic.  When $\sigma = 1$, the equations in \eqref{eq:6} are strictly hyperbolic
and genuinely nonlinear
\cite{kodama_whitham_1999}.

Modulation theory is a powerful tool for the analysis of multiscale
nonlinear waves in dispersive hydrodynamics
\cite{biondini_dispersive_2016}.  A prominent example is the
regularization of dispersive hydrodynamic singularities resulting in
dispersive shock waves (DSWs) that are expanding, modulated wavetrains
\cite{el_dispersive_2016}.  Utilizing the approach from
\cite{binswanger_whitham_2021} for the Whitham equation \eqref{eq:1},
we obtain the Whitham modulation equations for the FDNLS equation
\eqref{eq:generalized-NLS} and then apply them to the problem of
modulational instability.

\subsection{Classical MI of one- and two-phase solutions}
\label{sec:modul-inst-one}

The sign $\sigma$ in the NLS equation \eqref{eq:NLS-Classical}
determines the nature of the modulations as repulsive/defocusing
($\sigma = 1$) or attractive/focusing ($\sigma = -1$).
It will be helpful to review the stability of one- and two-phase
solutions of the defocusing and focusing NLS equation
\eqref{eq:NLS-Classical}.  Linearizing Eq.~\eqref{eq:NLS-Classical}
about the Stokes solution \eqref{eq:stokes_NLS}, and seeking
perturbations proportional to $e^{i((\overline{u}\pm k)x - (\gamma+\omega) t)}$ yields the linear dispersion relation 
\begin{align}
  \label{eq:2}
  (\omega - \ubar k )^2 = \frac{1}{4}k^2(k^2 + 4 \sigma \rhobar )~. 
\end{align}
For $k \to 0$, the phase speed
$\omega/k \to \ubar \pm \sqrt{\sigma \rhobar}$ limits to the
characteristic velocities of the modulation equations \eqref{eq:disperionless_modeq}.
The two branches of the frequency $\omega$ satisfying \eqref{eq:2},
are real-valued for all values of $k \in \mathbb{R}$ when
$\sigma = 1$.  When $\sigma = -1$, the growth rate of the positive
branch $\mathrm{Im} \, \omega = \tfrac12 |k| \sqrt{4 \rhobar - k^2}$
is nonzero for real $k$ satisfying $0 < k^2 < 4 \rhobar$.

To illustrate the instabilities of Stokes waves, we perform direct
numerical simulations of the NLS
equation~\eqref{eq:NLS-Classical}. The initial condition is the
perturbed Stokes wave $\psi(x,0) = e^{i\ubar x} + \epsilon p(x)$,
with $0 < \epsilon \ll 1$ and $p(x)$ smooth, band-limited noise,
\begin{align}
  \label{eq:noise_perturbation}
    p(x) &=  e^{i\ubar x} \sum\limits_{|n|< M}\mathcal{E}(n)e^{i 2 n \pi x /L}~, 
\end{align}
where $L$ is the computational domain size,
$M = \left\lfloor\frac{k L}{2\pi}\right\rfloor$ is the number of discrete Fourier
modes, and $\mathcal{E}(n)$ is a length $2M + 1$ vector with uniformly
sampled, random values in the interval $[-1,1]$ generated using
MATLAB's {\sc rand} function.
 
\begin{figure}[ht!]
  \centering
  \includegraphics[scale = 0.4]{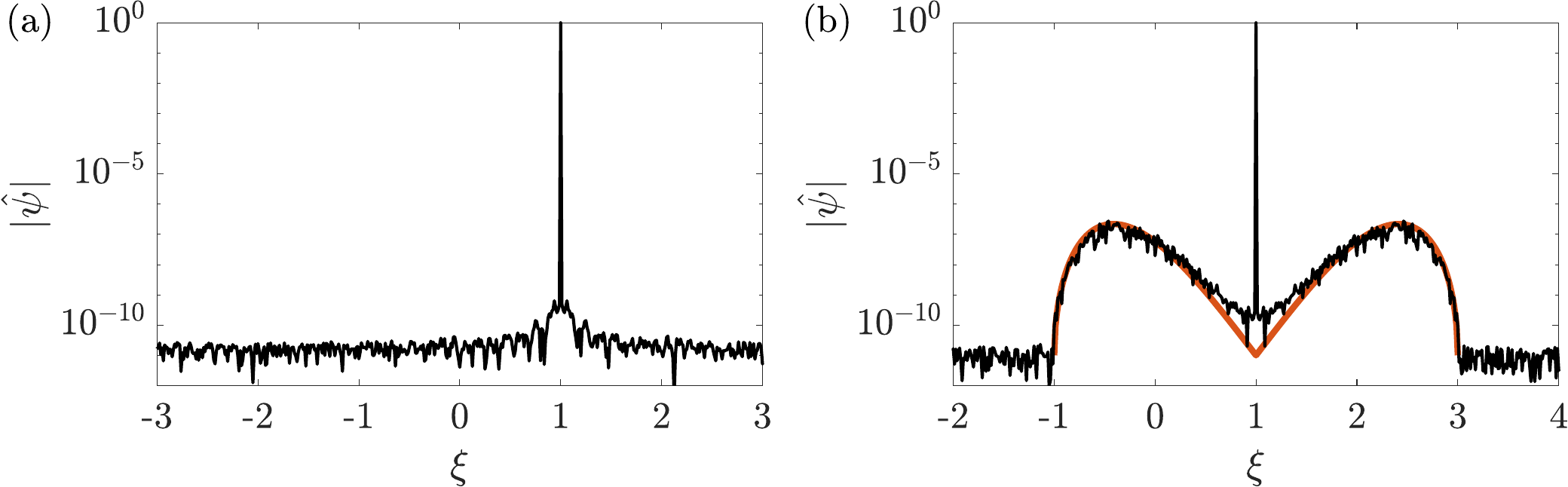}
  \caption{Spectral amplitude of the plane wave \eqref{eq:stokes_NLS}
    with $\rhobar = 1$, $\ubar = 1$ subject to the perturbation
    \eqref{eq:noise_perturbation} with $\epsilon = 10^{-11}$ after
    numerical evolution. (a) The defocusing case $\sigma = 1$ at
    $t = 20$, and $(b)$ the focusing case $\sigma = -1$ at $t = 10$
    with the predicted amplitude from linear theory in red.}
  \label{fig:MI_stokes_NLS}
\end{figure}

In Fig.~\ref{fig:MI_stokes_NLS}, we plot the amplitude of the Fourier
spectrum of the solution, $|\hat{\psi}(\xi,\Delta t)|$ with spectral
parameter $\xi$, after the evolution time $\Delta t > 0$.  In the case
$\sigma = 1$ of Fig.~\ref{fig:MI_stokes_NLS}(a), the spectrum consists
of a single peak at $\xi = \ubar$, with negligible change to the
spectrum of the initial perturbation of $\mathcal{O}(\epsilon)$.  For
$\sigma = -1$ in Fig.~\ref{fig:MI_stokes_NLS}(b), the spectrum for
$0 < |\xi - \overline{u}| < 2$ is amplified.  The amplitude is
predicted to grow according to eq.~\eqref{eq:2} (in which
$\xi = \ubar + k$) as
$|\hat{\psi}(\xi,\Delta t)| \approx |\hat{\psi}(\xi,0)| \exp \left (
  \mathrm{Im}\, \omega(\xi) \Delta t \right ) \approx \epsilon \exp
\left ( \tfrac12 |\xi - \ubar| \sqrt{4-(\xi-\ubar)^2} \Delta t \right
)$, provided $\Delta t$ is small enough for the evolution of the
perturbation \eqref{eq:noise_perturbation} to remain in the linear
regime. The predicted growth in the spectrum is overlaid on the
simulation in Fig.~\ref{fig:MI_stokes_NLS}(b).

Since the modulation equations \eqref{eq:disperionless_modeq} are
elliptic when $\sigma = -1$, ill-posedness of the initial value problem for the modulation equations corresponds to MI of the Stokes wave
\cite{whitham_linear_1974}.  Linearizing equations
\eqref{eq:disperionless_modeq} about constant $\rhobar$ and $\ubar$,
we observe that the growth rate of perturbations
$\propto e^{iK(X-\lambda_{1,2} T)}$ with wavenumber $K > 0$ is
$\mathrm{Im}\, \lambda_2 K = \sqrt{\rhobar} K$ when $\sigma = -1$.
This coincides with the small $k$ expansion of the growth rate from
the dispersion relation \eqref{eq:2},
$\mathrm{Im} \, \omega \sim \sqrt{\rhobar} k$, $k \to 0$.  Note that
modulation theory does not predict a saturation of the growth rate,
which, according to \eqref{eq:2}, occurs at the order one perturbation wavenumber
$k = \sqrt{2\rhobar}$.

The MI of the two-phase solution \eqref{eq:3} can be determined by the
hyperbolicity of the two-phase modulation equations \eqref{eq:6}.
Since the modulation equations are hyperbolic (elliptic) when
$\sigma = 1$ ($\sigma = -1$), the two-phase solution is modulationally
stable (unstable) and has been proven to be so by spectral analysis of
the linearized operator when $\sigma = 1$ \cite{bottman_elliptic_2011}
and $\sigma = -1$
\cite{gustafson_stability_2017,deconinck_stability_2017}. Furthermore,
it has been shown that weak hyperbolicity of the modulation equations
(all characteristic speeds are real) is a necessary condition for the
modulational stability of nonlinear periodic wave trains
\cite{bronski_modulational_2010,johnson_modulational_2020,benzoni-gavage_slow_2014}.
Sufficiency is obtained when the modulation equations are strictly
hyperbolic and the original PDE is Hamiltonian
\cite{johnson_m._a._rigorous_2010}.

\begin{figure}[H]
    \centering
    \includegraphics[scale = 0.8]{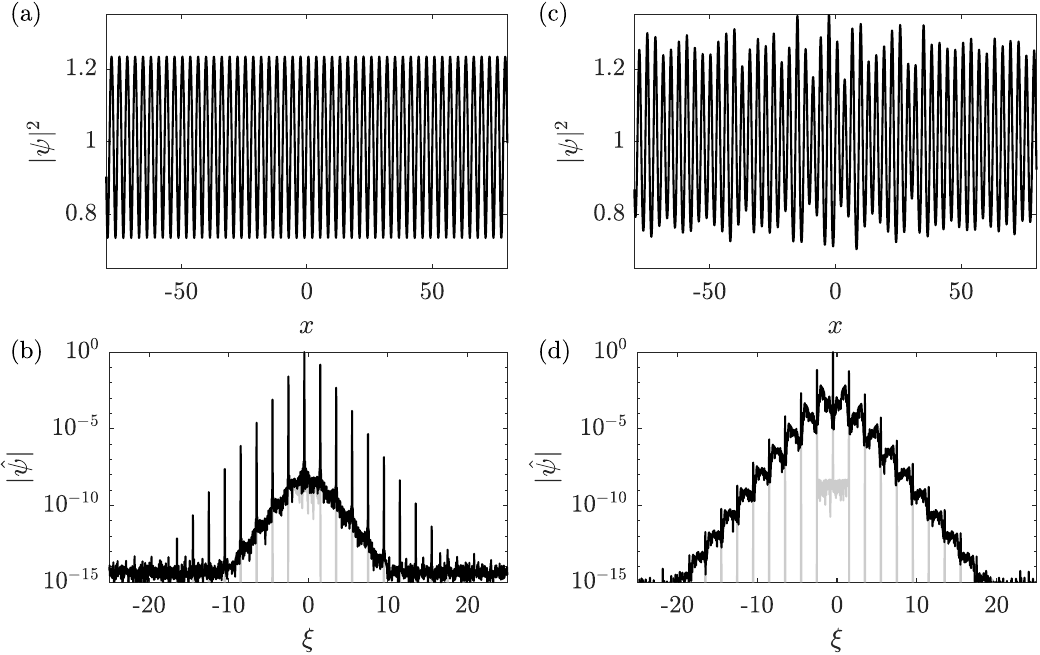}
    \caption{Initially perturbed two-phase solutions of the NLS
      equation with $\sigma = 1$ at $t = 150$ (a,b) and $\sigma = -1$ at $t = 10$ (c,d). 
      (a,c) depict the solution's square modulus and (b,d)
      show the Fourier amplitude spectrum at the initial (gray) and final
      (black) times.} 
    \label{fig:stability_NLS_examples}
\end{figure}

In Figure \ref{fig:stability_NLS_examples}, we plot two numerical
simulations of the perturbed two-phase solutions in the $\sigma = 1$
case (Fig.~\ref{fig:stability_NLS_examples}(a,b)) at $t = 150$ and the
$\sigma = -1$ case (Fig.~\ref{fig:stability_NLS_examples}(c,d)) at
$t = 10$, respectively.  Figures \ref{fig:stability_NLS_examples}(a,c)
depict the square modulus $|\psi|^2$ whereas
Figs.~\ref{fig:stability_NLS_examples}(b,d) show the amplitude
spectrum $|\hat{\psi}|$. For these simulations, the carrier wavenumber
is $\ubar = -0.5$ and the second phase wavenumber is $k = 2$, and the
amplitude parameter is $\rhobar = 1$. In both cases, there is a
prominent peak at $\xi = \ubar$ in the amplitude spectrum with
distinct harmonics at $\xi = \ubar + n k$, for integers $n$. In the
stable case, $\sigma = 1$, the spectrum is unchanged even for long
time evolution.

The spectral features in Fig.~\ref{fig:stability_NLS_examples}(d)
represent the early stage of MI, where the growth of the perturbation
has not exceeded the largest spectral harmonics. The nonlinear stage
of instability occurs for longer evolution times where the dynamics
can be quite complicated, though the integrability of the NLS equation
allows for the development of theory for both localized
\cite{biondini_universal_2016,biondini_long-time_2017} and random
\cite{gelash_bound_2019} perturbations.

In the simplest cases presented here, Eq.~\eqref{eq:NLS-Classical}
possesses both stable one-phase and two-phase solutions when
$\sigma = 1$ and both unstable one-phase and two-phase solutions when
$\sigma = -1$. This motivates the following question.  \emph{Does the
  addition of full dispersion into the NLS model
  \eqref{eq:generalized-NLS} allow for situations in which one-phase,
  plane wave solutions are stable, but two-phase solutions with the
  same $\rhobar$, $\ubar$ are unstable?}  After developing the
modulation analysis, we will provide several physically inspired
examples where the answer is yes.

\subsection{Outline of this manuscript}
This manuscript is organized as follows. In
Sec.~\ref{sec:modulation_theory}, we derive the Whitham modulation
equations for two-phase wave trains in the FDNLS equation
\eqref{eq:generalized-NLS}. In Sec.~\ref{sec:stability_gen}, we
utilize the Whitham modulation equations to identify criteria for when
both one-phase and two-phase wavetrains are stable with respect to
long wave perturbations. We also obtain approximate two-phase
solutions of the FDNLS equation with general $\Omega(k)$ for a plane
wave subject to weakly nonlinear modulations in the second
phase. Then, the corresponding two-phase Whitham modulation equations
and their characteristic velocities are obtained in this regime.  We
identify an index that determines the modulation equation type and the
different mechanisms that drive the instability of the underlying
two-phase solution. In Sec.~\ref{sec:examples}, we present phase
diagrams predicting the stable/unstable two-phase solutions and the
type of instability for several choices of dispersion $\Omega(k)$.
These predictions are then compared with numerical simulations of
perturbed two-phase solutions. Finally, in Sec.~\ref{sec:conclusions},
we conclude with a discussion of the implications and extensions of
the present theory.

\section{Modulation theory for two-phase waves}
\label{sec:modulation_theory}
 
The FDNLS equation \eqref{eq:generalized-NLS} admits the one-phase, Stokes
wave solutions
\begin{align}
\label{eq:theta1-gamma}
  \psi(x,t) = \sqrt{\rhobar} e^{i\theta_1}, \quad \theta_1 = \ubar x -
  \gamma t, \quad
  \gamma = \Omega(\ubar) + f'(\rhobar) ,
\end{align}
for any $\rhobar > 0$ and $\ubar \in \mathbb{R}$. Here, $\ubar$ and
$\gamma$ are the wavenumber, frequency pair, respectively.  The
solution \eqref{eq:theta1-gamma} is referred to as the carrier wave.
We assume the existence of a four-parameter family of two-phase
solutions to Eq.~\eqref{eq:generalized-NLS} in the form
\begin{align}
\label{eq:2phase_sol}
  \psi(x,t) = \mathcal{U}(\theta_2) e^{i\theta_1}, \quad 
  \theta_1 = \ubar x - \gamma t, \quad \theta_2 = k x - \omega t, 
\end{align}
where $(\ubar,\gamma)$ and $(k,\omega)$ are the first and second
phase's wavenumber and frequency, respectively.  The oscillation
periods for each phase are normalized to $2\pi$:
$\psi(\theta_1+2\pi n,\theta_2 + 2\pi m) = \psi(\theta_1,\theta_2)$
for integers $n$, $m$.  For incommensurate wavenumbers and
frequencies, the solution \eqref{eq:2phase_sol} is quasi-periodic.
Each of the frequencies $\gamma$ and $\omega$ generally depend on both
wavenumbers $\ubar$ and $k$.  The two-phase solution
\eqref{eq:2phase_sol} can be parameterized, for example, in terms of
the mean parameters
\begin{equation}
  \label{eq:8}
  \rhobar \equiv \frac{1}{4\pi^2} \int_0^{2\pi}\int_0^{2\pi}
  |\psi(\theta_1,\theta_2)|^2\,d\theta_1 d \theta_2, \quad
  \ubar \equiv \frac{i}{4\pi^2} \int_0^{2\pi}\int_0^{2\pi} (\psi_x^* \psi -
  \psi_x\psi^*)/|\psi|^2 \,d\theta_1 d \theta_2, 
\end{equation}
the wavenumber $k$, and amplitude $a$, which is the magnitude of the
Fourier mode with wavenumber $\ubar + k$.
The two-phase solution
\eqref{eq:2phase_sol} is an amplitude and phase modulated Stokes waves
\eqref{eq:theta1-gamma}.  We demonstrate existence by obtaining
approximate and numerical solutions in the weakly nonlinear regime.

Whitham modulation theory is a formal asymptotic procedure to derive a
system of conservation laws that describe the slow evolution of a
periodic or quasi-periodic solution
\cite{whitham_linear_1974,el_dispersive_2016}. We now carry out the
derivation of the modulation equations using Whitham's original method
of averaged conservation laws \cite{whitham_non-linear_1965}. A
modulated two-phase solution is sought in the form
\begin{align}
  \label{eq:modulation_ansatz}
  \psi(x,t) = \cU(\theta_2,X,T) e^{i\theta_1} + \epsilon
  \psi_1(\theta_1,\theta_2,X,T) + \cdots, \quad \epsilon \to 0,
\end{align}
where $\theta_1$ is the carrier phase, $\theta_2$ is the second,
envelope phase, and $\cU$ is a complex modulation, which depends on
$\theta_2$ and the slow scales $X = \epsilon x$, $T = \epsilon t$.
The phase functions $\theta_1,\theta_2$ are rapidly varying
\begin{align*}
  \theta_j = S_j(X,T)/\epsilon ~, \quad j=1,2,
\end{align*}
for smooth functions $S_1$, $S_2$.  It is expedient to define the
generalized wavenumbers and frequencies as
\begin{equation}
  \label{eq:phase_def}
  \begin{split}
    \theta_{1,x} &= S_{1,X} = \ubar, \qquad  \theta_{2,x} = S_{2,X} = k \\ 
    \theta_{1,t} &= S_{1,T} = -\gamma, \qquad  \theta_{2,t} = S_{2,T} = - \omega .
  \end{split}
\end{equation}
The periods of $S_1$ and $S_2$ are
fixed in order to ensure a well-ordered asymptotic expansion. Without
loss of generality, we take these to be $2\pi$, so that $\cU$ is also
$2\pi$-periodic in $\theta_2$. Hence, $\cU$ is represented by its
Fourier series
\begin{align}
\label{eq:Fourier}
\cU(\theta_2,X,T) = \sum_{n = -\infty}^{\infty} \hat{q}_n(X,T) e^{i n \theta_2}. 
\end{align}

It is a straightforward calculation to prove
that~\eqref{eq:generalized-NLS} possesses the following conserved
quantities 
\eqref{eq:2phase_sol}
\begin{subequations}
\label{eq:conserved_quantities}
\begin{align}
  \label{eq:mass}
  M[\psi] &=  \int |\psi|^2 \ dx , \\ 
  \label{eq:momentum}
  P[\psi] &= \int \left ( \psi^* \psi_x - \psi \psi^*_x  \right ) \ dx, \\ 
  \label{eq:energy}
  E[\psi] &= \int  \big ( \psi^* \Omega(-i\partial_x)\psi+ f(|\psi|^2)
  \big ) dx,
\end{align}
\end{subequations}
corresponding respectively to mass, momentum, and energy. 

Two modulation equations are found by inserting the multiple scales
expansion~\eqref{eq:modulation_ansatz} into the first two conservation
laws corresponding to $M$ and $P$ integrated over the two-phase
solution family.  Replacing partial derivatives in $x,t$ with
appropriate partial derivatives in $\theta_1, \theta_2, X$, and $T$,
the linear dispersion operator $\Omega$ is expanded in a similar way
to the expansions of the dispersion operators in the scalar Whitham
equation \cite{binswanger_whitham_2021}.  We extend those results to
the case where the operator is acting on a function of two independent
phase variables and obtain, to $\OO(\epsilon)$, that
\begin{align}
\begin{split}
\label{eq:op_expans}
\cO(-i\partial_x) \psi(x,t) &= 
\cO(\dd - i \epsilon \partial_X)
\psi(\theta_1,\theta_2,X,T)\\
& = \cO(D)\psi 
+ i \frac{\epsilon}{2} \left[\cO'(D)\psi_X 
+ \left(\cO'(D)\psi\right)_X\right] + \OO(\epsilon^2)~,
\end{split}
\end{align}
where $\dd = -i\partial_{\theta_1} - i \partial_{\theta_2}$, and
$\cO'(\cdot)$ is the pseudo-differential operator corresponding to the
symbol $\cOh'(\xi)$.  The proof for single-phase functions can be
found in the appendix of Ref.~\cite{binswanger_whitham_2021}, where
$\Omega(\xi)$ is assumed to be analytic. However, this requirement can
be relaxed to three weak derivatives of $\Omega(k)$ in an appropriate
function space \cite{clarke_rigorous_2022}.

We now outline the derivation of the modulation equation that is a
consequence of mass conservation \eqref{eq:mass}. First, one 
expands the conserved quantity~\eqref{eq:mass}
\begin{align}
  \begin{split}
    \label{eq:calc1}
    \frac{d}{dt}  \int_0^{2\pi}\!\!\!\int_0^{2\pi}  |\psi|^2 \ d\theta_1 d \theta_2 
     =\!\! \int_0^{2\pi}\!\!\!\int_0^{2\pi} (-\gamma \partial_{\theta_1} 
      - \omega \partial_{\theta_2} + \epsilon \partial_T) |\psi|^2 \
      d\theta_1 d\theta_2  = \epsilon \left(\overline{|\cU|^2} \right)_T  + \OO(\epsilon^2),
   \end{split}
\end{align}
where the averaging operator is denoted
\begin{align*}
\overline{F[\psi]}(X,T) =
  \int_0^{2\pi}\int_0^{2\pi}F[\psi(\theta_1,\theta_2,X,T)] \, d
  \theta_1 d \theta_2. 
\end{align*}
We now determine the representation of the averaged mass flux.  This
can be achieved by alternatively replacing $t$-derivatives of $\psi$
with the right hand side of Eq.~\eqref{eq:generalized-NLS} and
uses~\eqref{eq:Fourier} to find that
\begin{align}
  \begin{split}
    \label{eq:calc2}
    \frac{d}{dt}  \int_0^{2\pi}\int_0^{2\pi}  |\psi|^2 \ d\theta_1 d \theta_2 
      &= - \epsilon \left(\sum_n \Omega'(\ubar+nk) |q_n|^2 \right)_X +
        \OO(\epsilon^2).  
  \end{split}
\end{align}
Equating terms at $\OO(\epsilon)$ from \eqref{eq:calc1}
and~\eqref{eq:calc2} results in averaged mass conservation 
\begin{align}
  \label{eq:whitham_mass}
  \left(\overline{|\cU|^2}\right)_T + \left(\sum_n \Omega'(\ubar+nk)
  |q_n|^2 \right)_X = 0.
\end{align}

Similar calculations with \eqref{eq:momentum} and \eqref{eq:energy}, yield averaged
momentum conservation
\begin{align}
  \label{eq:whitham_momen}
  \begin{split}
    &\left(\sum_n (\ubar + nk)|q_n|^2\right)_T + \left(
      \overline{f'(|\cU|^2) |\cU|^2- f(|\cU|^2)}\right)_X  \\
    & \qquad \qquad \qquad \qquad ~\, + \left(\sum_n (\ubar+nk)
      \Omega'(\ubar + nk)|q_n|^2\right)_X = 0,
  \end{split}
\end{align}
and averaged energy conservation
\begin{equation}
  \label{eq:whitham_energy}
  \begin{split}
    &\left ( \overline{\cU^* \cO(D) \cU + f(|\cU|^2)} \right )_T +
      \left(\sum_n \Omega(\ubar + nk)  \Omega’(\ubar + nk)|q_n|^2\right)_X \\
  &\quad + \frac{1}{2}\left(\overline{\cU^*\left(\cO’(D) f’(|\cU|^2) \cU\right) +
     f’(|\cU|^2)\cU \left(\cO’(D)\cU^*\right)}\right)_X = 0 .
  \end{split}
\end{equation}

Two additional modulation equations are obtained by requiring that the
phase variables are twice continuously differentiable, \ie,
$S_{j,XT} = S_{j,TX}$ for $j = 1,2$.  These constraints result in the
two \textit{conservation of waves} equations
\begin{subequations}
    \begin{align}
\label{eq:cons_theta1}
\ubar_T + \gamma_X & = 0, \\ 
k_T + \omega_X & = 0.
\label{eq:cons_theta2} 
\end{align}
\end{subequations}

Equations \eqref{eq:whitham_mass}, \eqref{eq:whitham_momen},
\eqref{eq:whitham_energy}, \eqref{eq:cons_theta1}, and \eqref{eq:cons_theta2} are
five conservation laws for the four dependent variables
$(\rhobar,\ubar,k,a)$.  The averaged energy equation is, generically,
redundant with the remaining four modulation equations
\cite{whitham_linear_1974,benzoni-gavage_modulated_2021}. We will
focus on the averaged mass \eqref{eq:whitham_mass}, averaged momentum
\eqref{eq:whitham_momen}, and conservation of waves equations
\eqref{eq:cons_theta1}, \eqref{eq:cons_theta2} as a closed set of
modulation equations for the four modulation parameters
$(\rhobar,\ubar,k,a)$.  In what follows, we study properties of these
equations with increasing levels of complexity.

\section{Modulations of one- and two-phase wavetrains}
\label{sec:stability_gen}

\subsection{One-phase modulations}
\label{sec:dispersionless}

The modulation equations for the one-phase Stokes wave
$\psi = \sqrt{\rhobar}e^{i\theta_1}$ can be obtained from the
two-phase modulation equations by taking $q_n \to 0$ for $n \ne 0$
where $q_0 = \sqrt{\rhobar}$.  Then, Eqs.~\eqref{eq:whitham_mass} and
\eqref{eq:cons_theta1} become
\begin{subequations}
  \label{eq:dispersionless_whitham}
  \begin{align}
    \label{eq:dispersionless_mass}
    (\rhobar)_T + \left(\rhobar \Omega'(\ubar)\right)_X &= 0 \\*[2mm]
    \label{eq:dispersionless_cons_waves}
    \ubar_T + (f'(\rhobar) + \Omega(\ubar))_X & =0 ,
  \end{align}
\end{subequations}
When $q_n = 0$, $n \ne 0$, the momentum and energy conservation laws
\eqref{eq:whitham_momen}, \eqref{eq:whitham_energy} are an immediate
consequence of~\eqref{eq:dispersionless_whitham}.  The remaining
modulation equation \eqref{eq:cons_theta2} limits to
\begin{equation}
  \label{eq:linear_cons_waves}
  k_T + \left ( \omega_0^{(\pm)}  \right )_X = 0, 
\end{equation}
where $\omega_0^{(\pm)}$ is the linear dispersion relation of the
FDNLS equation \eqref{eq:generalized-NLS} for waves propagating on the
Stokes wave
\begin{equation}
  \label{eq:9}
  \begin{split}
    \omega_0^{(\pm)} &= \mathcal{N}_1 \pm
                       \sqrt{\mathcal{M}_1(\mathcal{M}_1 + 2 \rhobar
                       f''(\rhobar))}, \\
    \mathcal{N}_1 &= \tfrac12 (\Omega(\ubar + k) - \Omega(\ubar - k) )
                    , \quad \mathcal{M}_1 = \tfrac12\left(\Omega(\ubar + k) -
                    2\Omega(\ubar) + \Omega(\ubar - k)\right) .
  \end{split}
\end{equation}
The dispersionless modulation equations
\eqref{eq:dispersionless_whitham} are a $2\times 2$ system of
conservation laws for the variables $(\rhobar,\ubar)$ that are
independent of $k$.  We therefore first consider the evolution of
$(\rhobar,\ubar)$ according to \eqref{eq:dispersionless_whitham} and
then discuss the evolution of $k$ in \eqref{eq:linear_cons_waves}.

Equations \eqref{eq:dispersionless_whitham} exhibit the characteristic
velocities
\begin{align}
  \label{eq:classical_velocities}
  \lambda_{1,2} = \Omega'(\ubar) \pm \sqrt{\rhobar
  f''(\rhobar)\Omega''(\ubar)}~,
\end{align}
which are real and strictly ordered $\lambda_1 < \lambda_2$ if and
only if
\begin{align*}
    \rhobar f''(\rhobar) \Omega''(\ubar) > 0~.
\end{align*}
In this case, the modulation equations
\eqref{eq:dispersionless_whitham} are strictly hyperbolic.  Equations
\eqref{eq:dispersionless_whitham} are diagonalized in terms of
the Riemann invariants
\begin{align*}
  r_{1,2} = \int^{\ubar} \sqrt{\Omega''(\tau)} d\tau  
  \pm \int^{\rhobar} \sqrt{\frac{f''(s)}{s}}  ds ,
\end{align*}
so that Eqs.~\eqref{eq:dispersionless_whitham} are equivalent to
$r_{j,T} + \lambda_j r_{j,X} = 0$, $j=1,2$.  The system
\eqref{eq:dispersionless_whitham} is genuinely nonlinear so long as
$\nabla \lambda \cdot \mathbf{r} \ne 0$ for each
eigenvalue-eigenvector pair $(\lambda,\mathbf{r})$
\cite{lax_formation_1972}.  For cubic nonlinearity,
$f'(\rhobar) = \rhobar$, the loss of genuine nonlinearity occurs when
$3 \Omega''(\ubar)^{3/2} = \pm \sqrt{\rhobar} \Omega'''(\ubar)$.

The classical MI criterion
\begin{align}
\label{eq:classical-MI}
    \rhobar f''(\rhobar) \Omega''(\ubar) < 0~,
\end{align}
is obtained when the velocities \eqref{eq:classical_velocities} are
complex. We define the \textit{carrier-phase MI index}
\begin{align}
  \label{eq:classical-MI-index}
  \CPMI &= \sgn \left[ \rhobar f''(\rhobar) \Omega''(\ubar) \right]~,
\end{align}
so that the condition for classical MI is $\CPMI =-1$
\cite{whitham_linear_1974}.  This criterion is equivalent to
$\sigma = -1$ in the NLS equation \eqref{eq:NLS-Classical} for
modulations of a Stokes wave in dispersive, nonlinear media
\cite{binswanger_whitham_2021}.

The conservation of waves \eqref{eq:linear_cons_waves} describes the
evolution of infinitesimal (linear) waves propagating on a modulated
Stokes wave.  Its characteristic velocity is the group velocity
$\omega^{(\pm)}_{0,k}(k,\rhobar,\ubar)$ whose long wavelength limit
$\lim_{k\to 0} \omega^{(\pm)}_{0,k} = \lambda_{1,2}$ coincides with
one of Eq.~\eqref{eq:classical_velocities}.  When $\CPMI = -1$, the
instability's growth rate in the linear regime is
$\mathrm{Im} \,\omega^{(\pm)}_{0}$.  This linearization of the FDNLS
equation about the Stokes wave~\eqref{eq:theta1-gamma} and subsequent
analysis was previously identified as the extended criterion for
modulational instability \cite{amiranashvili_extended_2019}.  In what
follows, we investigate the scenario when the Stokes wave is
modulationally stable according to the extended MI criterion
$\CPMI = 1$ but exhibits an instability due to finite amplitude
modulations involving a second phase.




\subsection{Weakly nonlinear, two-phase wavetrains}
\label{sec:weakly_NL_lim}


To approximate the two-phase solution, we insert \eqref{eq:2phase_sol}
into~\eqref{eq:generalized-NLS} and
obtain
\begin{align}
\label{eq:per_PDE}
(-\gamma + i \omega \p{2})\cU & =  \cO\left(\ubar - i k \p{2}\right)\cU
\ +\ f'(|\cU|^2) \cU~. 
\end{align}
Introducing the small amplitude parameter $0 < a \ll 1$, expanding the
solution as
\begin{equation}
  \label{eq:Stokes-solution}
  \begin{split}
       \cU &= \sqrt{\rho_0}  + a 
    e^{i\theta_2} + a B_1 e^{-i\theta_2} + a^2 A_2 e^{2 i
          \theta_2} + a^2 B_2  e^{-2i\theta_2}+ \ldots , \\
    \gamma &= \gamma_0 \ +\ a^2 \gamma_2 \ +\ \ldots \qquad 
             \omega = \omega_0 \ +\ a^2 \omega_2 \ +\ \ldots ,
  \end{split}    
\end{equation}
inserting the expansions into~\eqref{eq:per_PDE}, and gathering terms
in powers of $a$, we solve the resulting linear problems for $A_2$,
$B_j$, $\gamma_j$, $\omega_j$ for $j = 1,2$.  The results of the
perturbation analysis up to $\OO(a^2)$ are
\begin{subequations}
  \label{eq:stokes_wave_coeffs}
  \begin{equation}
    \label{eq:primary-vars}
    \begin{split}
      \gamma_0 &= f'(\rho_0) + \Omega(\ubar), \quad \omega_0^{(\pm)} = \N_1
                 \pm \sqrt{\PP}, \quad B_1^{(\pm)}  = -1 + \frac{\M_1 \pm
                 \sqrt{\PP} }{\rho_0 f''(\rho_0)}, \\*[2mm]
      A_2^{(\pm)} & = \frac{\sqrt{\rho_0}}{\D}\left[(\M_2-\N_2 + 2
            \gamma_0)\left( \FF_3 + 2(2 B_1^{(\pm)} +1) f''(\rho_0)\right)
            - \FF_2 \right] ,\\*[2mm] 
      B_2^{(\pm)} & = -\frac{\sqrt{\rho_0}}{\D}\left[(\M_2+\N_2-2 \gamma_0)
            \left( \FF_3 + 2 B_1^{(\pm)}  (B_1^{(\pm)} +2) f''(\rho_0)\right) -
            \FF_2 \right], \\*[2mm] 
      \gamma_2^{(\pm)} & = 2 \left ( \left ( B_1^{(\pm)} \right )^2 +
                         B_1^{(\pm)} + 1 \right )f''(\rho_0) + \FF_3,
    \end{split}
  \end{equation}
  and
  {\footnotesize
    \begin{equation}
      \label{eq:omega2_gen}
      \begin{split}
        \omega_2^{(\pm)}
        & =  \frac{1}{2((B_1^{(\pm)})^2-1)} \bigg[2 \left((1 +
          2B_1^{(\pm)})A_2^{(\pm)} + B_1^{(\pm)} (2 + B_1^{(\pm)})
          B_2^{(\pm)}\right) \\
        & \quad \times \left((B_1^{(\pm)})^4 + 2 (B_1^{(\pm)})^3 +
          2 B_1^{(\pm)} + 1 - 2 \sqrt{\rho_0} f''(\rho_0) \right)  \\ 
        & \quad - \left(\sqrt{\rho_0}(A_2^{(\pm)} + B_2^{(\pm)}) + (1 +
          (B_1^{(\pm)})^2)\right)\FF_3+ \rho_0 f^{(4)}(\rho_0)(1 +
          B_1^{(\pm)})^4 \bigg],
      \end{split}
    \end{equation}}
  where we introduce
  \begin{align}
    \label{eq:bonus-variables}
      \M_{j} &= \tfrac12\left(\Omega(\ubar + jk) - 2\Omega(\ubar) +
               \Omega(\ubar -jk)\right), \quad 
               \N_j = \tfrac12\left(\Omega(\ubar+jk) - \Omega(\ubar -
               jk)\right), \nonumber \\
      \PP & = \M_1\left(\M_1 + 2 \rho_0 f''(\rho_0)\right)~, \quad
            \FF_2 = 2 \left(B_1 ^2-1\right) \rho_0 [f''(\rho_0)]^2, 
      \\
      \FF_3 &= (B_1 +1)^2 \rho_0 f^{(3)}(\rho_0), \quad 
              \D = 2 (2\gamma_0 - \N_2)^2 - \M_2\left(\M_2 + 2\rho_0
              f''(\rho_0) \right) .\nonumber
  \end{align}
\end{subequations}
The approximate two-phase solution is given
by~\eqref{eq:Stokes-solution} and \eqref{eq:stokes_wave_coeffs}. A few
remarks are in order.  
There are two solutions for $\omega_0^{(\pm)}$ as in
\eqref{eq:9}. Hence, there are two distinct branches of two-phase
solutions bifurcating from the one-phase solution. The denominator of
$\omega_2^{(\pm)}$ in \eqref{eq:omega2_gen} is zero when
$|B_1^{(\pm)}|= 1$.  This occurs only when $\M_1=0$ or $\rho_0=0$,
which we exclude from further consideration.  In general, $\M_j$ and
$\N_j$ depend on both $\ubar$ and $k$, and hence so do
$\omega_0^{(\pm)},\gamma_2$, and $\omega_2^{(\pm)}$.  The terms $\M_j$
and $\N_j$ are the discrete Laplacian and centered difference,
respectively, that capture the dispersion's nonlocality with the long
wavelength limits $\M_j \sim \frac{1}{2}(jk)^2\cOh''(\ubar)$,
$\N_j \sim \frac{1}{2}jk\cOh'(\ubar)$ as $k \to 0$.

     
\subsection{Modulation system for weakly nonlinear two-phase
  solutions}

Inserting the two-phase solution
\eqref{eq:Stokes-solution}--\eqref{eq:stokes_wave_coeffs} into the
modulation system for mass \eqref{eq:whitham_mass}, momentum
\eqref{eq:whitham_momen}, and conservation of waves
\eqref{eq:cons_theta1}, \eqref{eq:cons_theta2} while retaining terms
up to $\OO(a^2)$, we arrive at
\begin{subequations}
  \label{eq:weakly_NL_whitham2}
  \begin{align}
    \label{eq:7}
    \left(\rho_0 + a^2(B_1^2 + 1)\right)_T  
    + \left(\rho_0 \Omega'(\ubar) + a^2 \QQ_+ \right)_X &= 0~, \\*[2mm]
    \begin{split}
      \left(\rho_0\ubar + a^2( \ubar+k + B_1^2(\ubar-k)) \right)_T
      \hspace{1.9in}
      \\*[2mm]
      + \left[  - f(\rho_0) + \rho_0 \left(f'(\rho_0) + \ubar
          \Omega'(\ubar) \right) + a^2\left(\rho_0  \gamma_2  + 
          \ubar\QQ_+ + k\QQ_- \right)\right]_X
      & = 0~,
    \end{split}
    \\*[2mm]
    \label{eq:10}
    \ubar_T + (\gamma_0 + a^2 \gamma_2)_X &= 0~, \\*[2mm]
    \label{eq:11}
    k_T + (\omega_0 + a^2 \omega_2)_X  &= 0~,
  \end{align}
\end{subequations}
where we introduce
\begin{align}
\label{eq:Q}
    \QQ_\pm &= \Omega'(\ubar + k) \pm  B_1^2\Omega'(\ubar-k),
\end{align}
and have suppressed the superscript $^{(\pm)}$ denoting the fast
($+$)/slow ($-$) branch of two-phase solutions---which differs from the
subscript in $\QQ_\pm$---for ease of presentation.  Note that $\QQ_\pm$
depend also on $\rho_0$ via $B_1$ \eqref{eq:primary-vars}.

\subsection{Characteristic velocities of the weakly nonlinear
  modulation system}
\label{sec:characteristic_vels}

To compute the characteristic velocities $\{\lambda_j\}_{j=1}^4$ of
the  Whitham modulation equations
\eqref{eq:weakly_NL_whitham2}, we can cast them in the quasilinear
form
\begin{align}
  \mathcal{A} \mathbf{q}_T + \mathcal{B} \mathbf{q}_X = 0, \qquad \mathbf{q} = \left[\rho_0, \ubar, a^2,k\right]^{\rm T} \end{align}
and solve the generalized eigenvalue problem  $\mathcal{B} \mathbf{v}
= \lambda \mathcal{A} \mathbf{v}$. We compute the eigenvalues perturbatively in the amplitude parameter $a$. At leading order, the eigenvalues are 
\begin{align}
  \label{eq:lambda12}
  \lambda_{1,2} & = \Omega'(\ubar) \pm \sqrt{\rho_0 f''(\rho_0)\Omega''(\ubar)}~,\\*[2mm]
  \label{eq:lambda34}
  \lambda_{3} & = \lambda_4 = \frac{\partial \omega_0}{\partial k}~. 
\end{align}
Here, $\lambda_{1,2}$ are simple eigenvalues for the mean variables
$\rhobar = \rho_0 + \OO(a^2)$ and $\ubar$ (the same as in
\eqref{eq:classical_velocities}).  The double eigenvalue is
degenerate, with geometric multiplicity one. Then, $\OO(a^2)$
perturbation of the eigenvalue problem will lead to $\OO(a^2)$
corrections to $\lambda_{1,2}$ and an $\OO(a)$ bifurcation of the
double eigenvalue $\lambda_3 = \lambda_4$. We are interested in the
$\OO(a)$ bifurcation of the double eigenvalue.  Consequently, we can
avoid a full perturbative analysis of the $4\times 4$ eigenvalue
problem in favor of a simpler, more direct approach that utilizes the
structure of the modulation equations themselves. That is, we assume
that the variations in the mean $(\rhobar,\ubar)$ are induced solely
by the finite amplitude wave while the leading order mean is taken to
be constant \cite{whitham_linear_1974,el_unsteady_2006}.  We make the
ansatz
\begin{align}
  \label{eq:induced-means}
  \rho_0 &= R_0 + a^2 R_2(k), \quad
           \ubar = U_0 + a^2 U_2(k),
\end{align}
where $(R_0,U_0)$ is the constant background and $(R_2,U_2)$ is the
induced mean.  The latter can be determined by
inserting~\eqref{eq:induced-means} into~\eqref{eq:7} and
\eqref{eq:10}, and recalling~\eqref{eq:stokes_wave_coeffs}
and~\eqref{eq:Q}.  The leading order terms vanish, while the
$\OO(a^2)$ terms yield
\begin{subequations}
\label{eq:induced_mean_mod}
\begin{align}
\left( R_2  +1 + B_1^2 \right)_T 
+ \left(  \Omega'(U_0) R_2  + \Omega''(U_0)R_0 U_2 + \QQ_{+,0} \right)_X & = 0~, \\*[2mm]
  U_{2,T}  + 
\left(  f''(R_0)R_2 +   \Omega'(U_0)U_2 +  \gamma_2 \right)_X & = 0~,
\end{align}
\end{subequations}
where $\QQ_{+,0} = \QQ_+(R_0,U_0)$.
 
As a consequence of the conservation of waves \eqref{eq:cons_theta2},
for any differentiable function $F(k(X,T))$, the conservation law
\begin{align}
\label{eq:conservation_form}
\left( F(k)\right)_T + \left(   \omega_{0,k} F(k)\right)_X = 0 
\end{align}
is satisfied. 
Comparing~\eqref{eq:induced-means}--\eqref{eq:conservation_form} with
$F = R_2 + 1 + B$ and $F = U_2$, we arrive at the algebraic system for
the induced mean
\begin{align*}
  \begin{bmatrix}
    \Omega'(U_0) - \omega_{0,k} & \Omega''(U_0) R_0 \\
    f''(R_0) & \Omega'(U_0) - \omega_{0,k}
  \end{bmatrix}
  \begin{bmatrix}
    R_2 \\ U_2
  \end{bmatrix}
  =
  \begin{bmatrix}
    \omega_{0,k}(1+B_1^2) - \mathcal{Q}_{+,0} \\ - \gamma_2    
  \end{bmatrix},
\end{align*}
whose solution is
\begin{align}
  \label{eq:induced_means_solution}
  \begin{bmatrix}
    R_2 \\ 
    U_2 
  \end{bmatrix}
  & = \frac{1}{\Delta}
    \begin{bmatrix}
      \left(\omega_{0,k} - \Omega'(U_0)\right) 
      \left( \QQ_{+,0} - (1+B_1^2)  \right) +
      \Omega''(U_0) R_0 \gamma_2 \\*[2mm] 
      \left(\omega_{0,k} - \Omega'(U_0)\right) \gamma_2 
      - f''(R_0)\left(\QQ_{+,0} - (1+B_1^2) \right) 
    \end{bmatrix},
\end{align}
with determinant 
\begin{align}
  \label{eq:13}
  \Delta &= \left(\omega_{0,k} -  \Omega'(U_0)\right)^2 -
           R_0f''(R_0)\Omega''(U_0)~. 
\end{align}
The nonlinear frequency shift is obtained by expanding
\begin{align*}
  \omega_0(R_0 + a^2 R_2, U_0 + a^2 U_2)
  &= \omega_0(R_0,U_0) + a^2 \tilde{\omega}_2  + \OO(a^4),  
\end{align*}
where
\begin{align}
  \label{eq:omega2-tilde}
  \tilde{\omega}_2(k) &= \omega_2(R_0,U_0) +
                        \left[R_2(k),  U_2(k)\right] \cdot
                        \left.\nabla_{\rho_0,\ubar}\omega_0\right|_{\rho_0 
                        = R_0,\ubar = U_0} ,
\end{align}
and $\omega_2$ is given in eq.~\eqref{eq:omega2_gen}.
Incorporating the nonlinear frequency shift \eqref{eq:omega2-tilde}
into the modulation equations \eqref{eq:7} and \eqref{eq:11} results
in the modulation system
\begin{equation}
  \label{eq:12}
  a_T + \omega_{0,k} a_X + \tfrac12 \omega_{0,kk} a k_X = 0, \quad k_T
  + \omega_{0,k} k_X + 2 a \tilde{\omega}_2(k) a_X = 0 .
\end{equation}
The characteristic velocities of \eqref{eq:12} include the sought for
perturbations of the degenerate, double eigenvalue
\eqref{eq:lambda34}.  Then, the characteristic velocities of the
weakly nonlinear modulation system \eqref{eq:weakly_NL_whitham2} are
\begin{subequations}
  \label{eq:whitham_velocities}
  \begin{align}
    \lambda_{1,2} & = \Omega'(\ubar) \pm \sqrt{\rho_0 f''(\rho_0)
                    \Omega''(\ubar)} + \OO(a^2)~, \\*[2mm]
    \lambda_{3,4} & = \omega_{0,k} \pm a
                    \sqrt{\omega_{0,kk}\tilde{\omega}_2} + \OO(a^2)~.
  \end{align}
\end{subequations}

\subsection{Generalized MI for small amplitude two-phase wavetrains}
\label{sec:Generalized-MI}

For the basis of this discussion, we assume that the dispersionless
system is hyperbolic, \ie, $\lambda_{1,2}$ are real.  We further
assume that the two-phase wavetrain, in the harmonic limit does not
experience exponential growth or decay, meaning that $\omega_0$ is
real valued.  Under these assumptions, the only mechanism that remains
to change the type of the modulation equations is a change of the
second-phase MI index
\begin{align}
\label{eq:MI_index_second_phase}
\SPMI = \sgn \left ( \omega_{0,kk} \tilde{\omega}_2 \right )~. 
\end{align}
This index provides a generalized criterion for MI of two-phase waves.
When $\SPMI=1$, the characteristic velocities $\lambda_{3,4}$ are
real.  When $\CPMI=\SPMI=1$, the  modulation system
\eqref{eq:weakly_NL_whitham2} is hyperbolic.  When $\SPMI=-1$,
$\lambda_{3,4}$ are complex valued and, if $\CPMI=1$, the modulation
equations are of mixed hyperbolic-elliptic type.

By examining $\SPMI$, we can identify potential mechanisms for a
change in type of the modulation system when
$\omega_{0,kk} \tilde{\omega}_2$ is zero or singular.  These include:
\begin{itemize}
\item zero linear dispersion curvature: $\omega_{0,kk} = 0$;
\item four-wave mixing: $\omega_{0,kk} \to \infty$ when
  $\Omega(\ubar + 2k) - 2 \Omega(\ubar) + \Omega(\ubar - 2k) = 0$
  ($\M_2=0$);
\item short-long wave resonance: $\Delta = 0$ in
  \eqref{eq:induced_means_solution} and \eqref{eq:13} when
  $\omega_{0,k} = \Omega'(\ubar) \pm \sqrt{\rhobar_0
    f''(\rhobar)\Omega''(\ubar)}$;
\item second harmonic resonance: $A_2$, $B_2 \to \infty$ in
  \eqref{eq:Stokes-solution} when
  $2 \omega_0(k,\rhobar,\ubar) = \omega_0(2k,\rhobar,\ubar)$ ($\D=0$ in
  \eqref{eq:bonus-variables});
 \item other nonlinear mechanisms: $\tilde{\omega}_2 = 0$.
\end{itemize}

\section{Examples}
\label{sec:examples}

In this section, we analyze the second-phase index $\SPMI$
\eqref{eq:MI_index_second_phase} for some specific FDNLS equations as
a way to predict the onset of instability. The FDNLS equation
\eqref{eq:generalized-NLS} is considered with cubic nonlinearity
$f'(|\psi|^2) = |\psi|^2$ and four distinct dispersion relations
$\Omega(k)$ corresponding to third and fourth-order dispersion as well
as finite-depth water waves and discrete nonlinear lattices.  Without
loss of generality, we consider normalized two-phase solutions
\eqref{eq:2phase_sol} with $\rhobar = \overline{|\mathcal{U}|^2} = 1$.
Since our objective is to identify instabilities using $\SPMI$, we
only consider modulationally stable one-phase carrier wavenumbers
$\ubar$ for which $\CPMI = 1$ \eqref{eq:classical-MI-index}.

In Secs.~\ref{sec:NLS3} and \ref{sec:NLS4}, we numerically compute
two-phase solutions of the FDNLS equation \eqref{eq:generalized-NLS}
of the form \eqref{eq:2phase_sol} by solving the nonlinear eigenvalue
problem
\begin{equation}
  \label{eq:NL_eigenvalue_U}
  -i\omega \mathcal{U} + \gamma \mathcal{U}_{\theta_2} =
  |\mathcal{U}|^2 \mathcal{U} + \Omega(\ubar -
  ik\partial_{\theta_2})\mathcal{U} ,
\end{equation}
using an iterative Newton-conjugate gradient algorithm for the Fourier
coefficients of $\mathcal{U}$ \cite{yang2009newton,yang2010nonlinear},
initialized with the weakly nonlinear
approximation~\eqref{eq:Stokes-solution}. Direct simulations of the
FDNLS equation \eqref{eq:generalized-NLS} are then performed using a
fourth-order split-step scheme~\cite{yoshida1990construction} with
initial condition the two-phase solution perturbed with small additive
noise \eqref{eq:noise_perturbation} with wavenumbers in the band
$\xi \in (\ubar-k,\ubar+k)$, similar to the numerical simulations of
perturbed solutions in Sec.~\ref{sec:modul-inst-one}.

\subsection{NLS with third-order dispersion}
\label{sec:NLS3}

\begin{figure}[H]
    \centering
    \includegraphics[scale = 0.42]{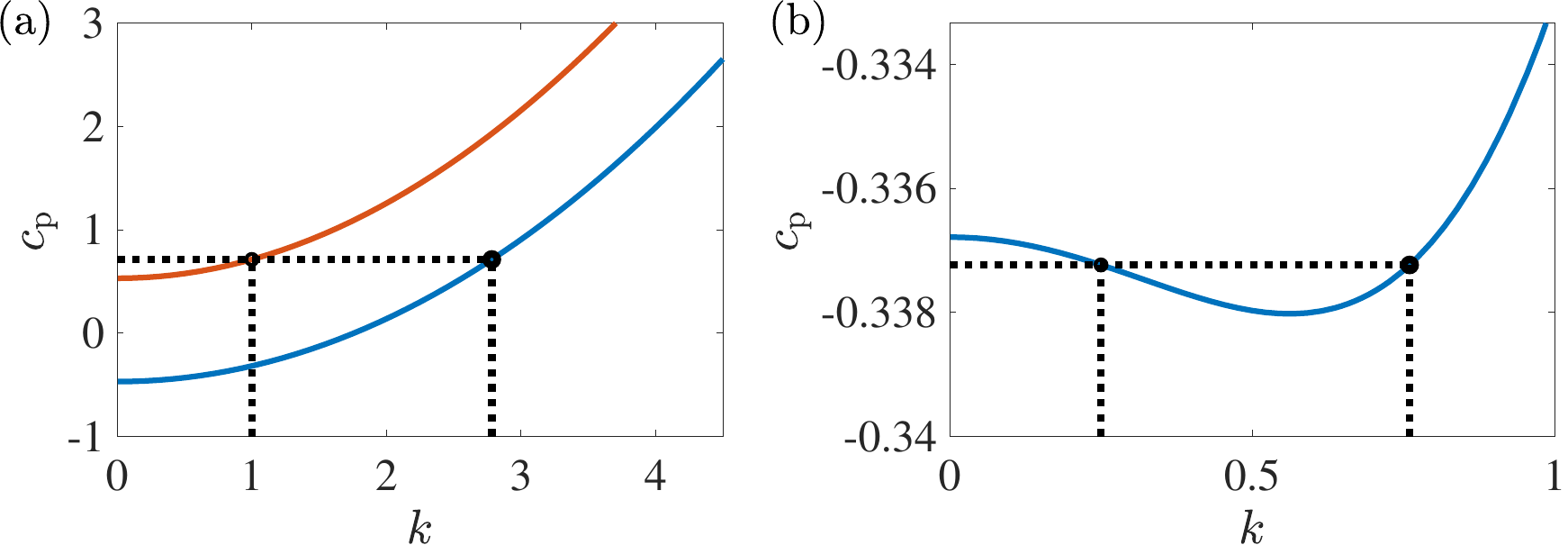}
    \caption{Example linear resonances for the NLS3
      equation~\eqref{eq:NLS3}: (a) inter-branch resonance
      $c_+(k_1) = c_-(k_2)$ of the fast $k_1 = 1$ (red) and slow
      $k_2 \approx 2.789$ (blue) dispersion branches for
      $\ubar = 0.25$; (b) intra-branch resonance $c_-(k_1) = c_-(k_2)$
      for $k_1 = 0.25$, $k_2 \approx 0.76$, $\ubar = 1.5$.}
    \label{fig:linear_resonances}
\end{figure}

The cubic NLS equation with third-order dispersion (NLS3) is
\begin{align}
\label{eq:NLS3}
  i \psi_t = \frac{i}{6}\psi_{xxx} + |\psi|^2\psi .
\end{align}
The linear dispersion relation is
\begin{align}
  \left ( \omega_0 - \frac{1}{2}\ubar^2 k - \frac{k^3}{6} \right )^2 =
  \ubar k^2\left(4 + \ubar k^2\right)~. 
\end{align}
When $\ubar \geq 0\,$, $\omega_0^{(\pm)} \in \mathbb{R}$ so that the
one-phase solution \eqref{eq:theta1-gamma} is modulationally stable.
The two branches of the dispersion relation correspond to the phase
velocities $c_\pm(k) = \omega_0^{(\pm)}(k)/k$ subject to inter-branch
($c_+(k_1) = c_-(k_2)$) and intra-branch ($c_+(k_1) = c_+(k_2)$) linear
resonances.  Despite being linear resonances, they can be induced by
nonlinearity, as demonstrated below.  Figures
\ref{fig:linear_resonances}(a) and \ref{fig:linear_resonances}(b)
depict inter- and intra-branch resonances, respectively.

\begin{figure}[H]
  \centering
  \includegraphics[width=0.99\textwidth]{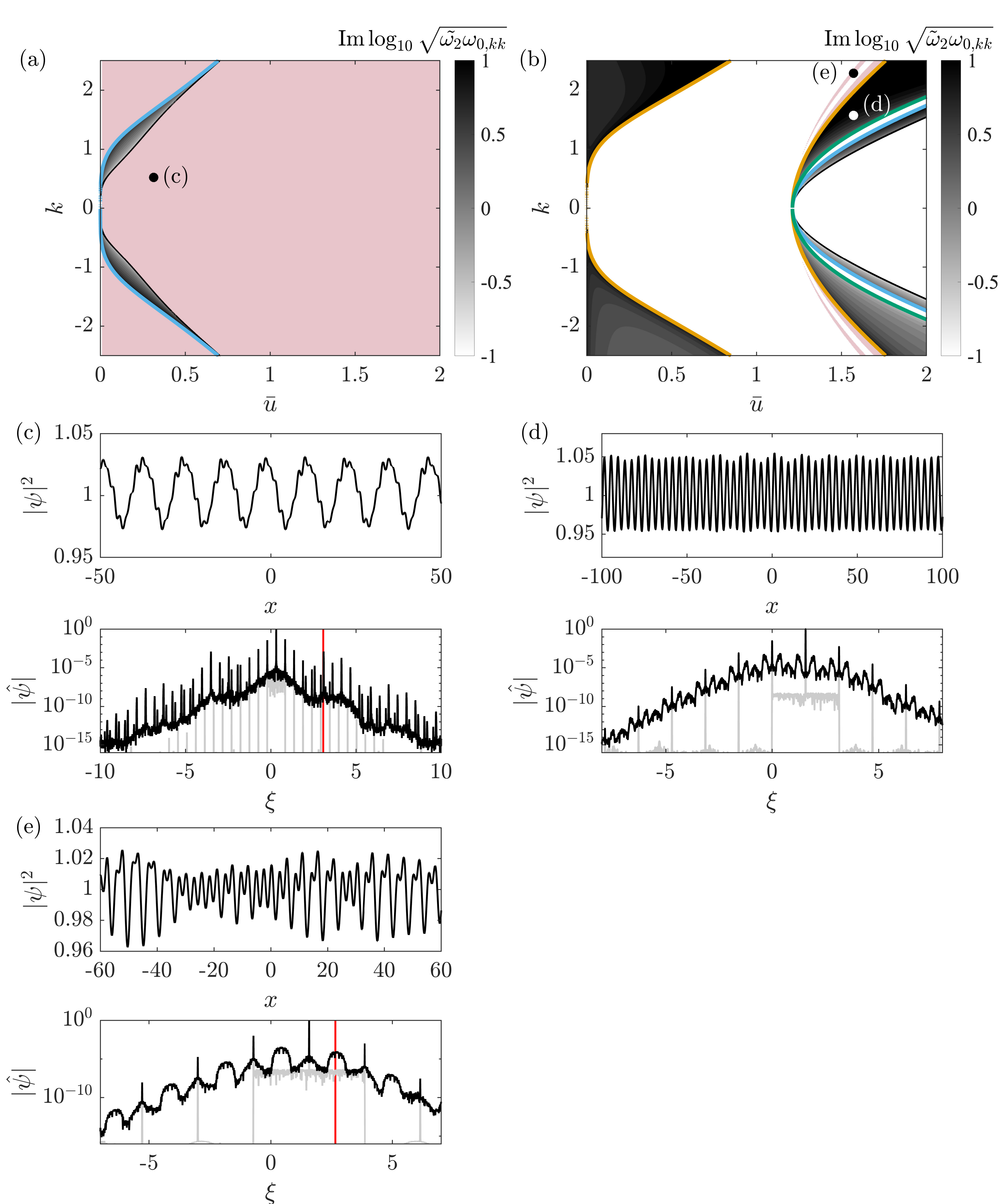}
  \caption{
    Phase diagrams indicating the stability (white) or instability
    (grayscale) of small-amplitude, fast (a) and slow (b) two-phase
    solutions of the NLS3 equation~\eqref{eq:NLS3}. The grayscale
    indicates the predicted strength of the instability. Pink regions
    indicate the presence of an unstable resonant mode. (c)-(e)
    present the results of direct numerical simulations of the NLS3
    equation. The upper (lower) panels show the power $|\psi(x,t)|^2$
    (spectral intensity $|\hat{\psi}(k,t)|$) in black after time
    integration. The initial spectrum $|\hat{\psi}(k,0)|$ is shown in
    gray. Red lines identify $\ubar + k_{\rm res}$.}
    \label{fig:NLS3_examples}
\end{figure}

We evaluate $\SPMI$ \eqref{eq:MI_index_second_phase} for
Eq.~\eqref{eq:NLS3} and plot the results in
Fig.~\ref{fig:NLS3_examples}(a,b).  The grayscale colormap depicts the
imaginary part of the Whitham velocities~\eqref{eq:whitham_velocities}
$\mathrm{Im}(\sqrt{\tilde{\omega_2} \omega_{0,kk}})$ on a log scale,
taken as a quantitative measure of the strength of the instability, on
the order of the exponential growth rate.  Figures
\ref{fig:NLS3_examples}(a) and (b) depict $\SPMI$ for the fast
$^{(+)}$ and slow $^{(-)}$ branches, respectively, of two-phase
solutions \eqref{eq:Stokes-solution} as a function of $\ubar$ and $k$.
Grayscale regions correspond to negative index $\SPMI = -1$ and
unstable two-phase solutions.  White regions correspond to positive
index $\SPMI = 1$.  The pink regions in
Fig.~\ref{fig:NLS3_examples}(a) indicate the existence of either an
inter- or intra-branch resonant wave with $\SPMI = -1$.


In Figures \ref{fig:NLS3_examples}(a,b),
\ref{fig:NLS_4th_examples}(a,b), \ref{fig:stability_waterwaves}, and
\ref{fig:stability_dnls} depicting $\SPMI$, the solid curves identify
one of the instability mechanisms listed in
Sec.~\ref{sec:Generalized-MI}. An orange curve identifies a short-long
wave resonance. Blue denotes a second-harmonic resonance. Green indicates zero-dispersion curvature whereas
black connotes other nonlinear mechanisms.  In this example there are no changes in sign of the $\SPMI$ due to 4-wave mixing. Figure~\ref{fig:NLS3_examples}(a) predicts
that the fast two-phase solution \eqref{eq:Stokes-solution} is
unstable for all wavenumber pairs $(\ubar,k)$ due to either
second-harmonic resonance, other nonlinear mechanisms, or the
secondary instability of a linearly resonant mode.  Figure
\ref{fig:NLS3_examples}(b) predicts bands of unstable, slow two-phase
solutions due to all available mechanisms.

To test the accuracy of our predictions, we compare them to the
numerical evolution of Eq.~\eqref{eq:NLS3} for several computed
two-phase solutions.  Figure \ref{fig:NLS3_examples}(c) shows the wave
power $|\psi(x,t)|^2$ and magnitude of the Fourier transform
$|\hat \psi(\xi,t)|$ at time $t = 2500$ corresponding to the
perturbed, fast two-phase solution with $(\ubar,k) = (\pi/10,\pi/3)$
and $a = 0.05$ identified in Fig.~\ref{fig:NLS3_examples}(a). For
these parameters, there is an inter-branch resonant mode with
wavenumber $k_{\rm res} \approx 2.789$.  Since the slow two-phase
solution with $(\ubar,k) = (\pi/10,k_{\rm res})$ is predicted to be
unstable, we can interpret the small amplitude, shorter wavelength
modulations of $|\psi|^2$ as caused by the slow growth of a resonant
mode. Indeed, the Fourier spectrum at $t = 2500$ shows a peak at
$k \approx \ubar + k_{\rm res}$ indicated by the red line that was not
in the initial data (gray).

Figure \ref{fig:NLS3_examples}(d) shows an example of an unstable mode at $t =1500$. In the long time evolution, modulations of $|\psi|^2$ are visible in
Fig.~\ref{fig:NLS3_examples}(d) due to the two-phase instability. The
spectrum reveals the emergent side-bands, which are indicated by the
peaks at $\ubar + n k$ for $n \in \mathbb{Z}$.

Figure \ref{fig:NLS3_examples}(e) is an example of an instability
arising due to a longer wave intra-branch resonance. We evolve a
perturbed, slow two-phase solution with $(\ubar,k) = (\pi/2,8\pi/11)$
and $a = 0.001$ to $t = 1000$. There is an intra-branch resonant mode
at $k_{\rm res} \approx 1.078$. In this case, the initial two-phase
solution is predicted to be stable, while the resonant mode is
predicted to be unstable. The power in Fig.~\ref{fig:NLS3_examples}(e)
exhibits strong modulations with wavelength longer than
$2\pi/k = 11/4$. The spectrum reveals wide bands centered at
$\ubar + k_{\rm res}$ and its harmonics. These bands' ampltiude grows
with time.  Despite this growth, the spectral peaks of the initial
two-phase solution remain relatively unchanged on the time scale of
the numerical experiment.

\subsection{NLS with fourth-order dispersion}
\label{sec:NLS4}

Consider the cubic NLS equation with fourth-order dispersion (NLS4)
\begin{align}
\label{eq:NLS4}
    i \psi_t = \frac{1}{24}\psi_{xxxx} + |\psi|^2\psi ~.
\end{align}
This equation exhibits no classical MI since $\CPMI = 1$ in
\eqref{eq:classical-MI-index}.  The linear dispersion relation
$\omega_0$ satisfies
\begin{align}
    \left(6 \omega_0 - \ubar k (\ubar^2 + k^2)\right)^2=\frac{1}{16}
  k^2 \left(k^2+6 \ubar^2\right) \left(k^4+6 k^2 \ubar^2+48 \right) ,
\end{align}
so that $\omega_0$ is real-valued for all $\ubar,k \in \mathbb{R}$.

\begin{figure}[h!]
  \centering
  \includegraphics[width=0.99\textwidth]{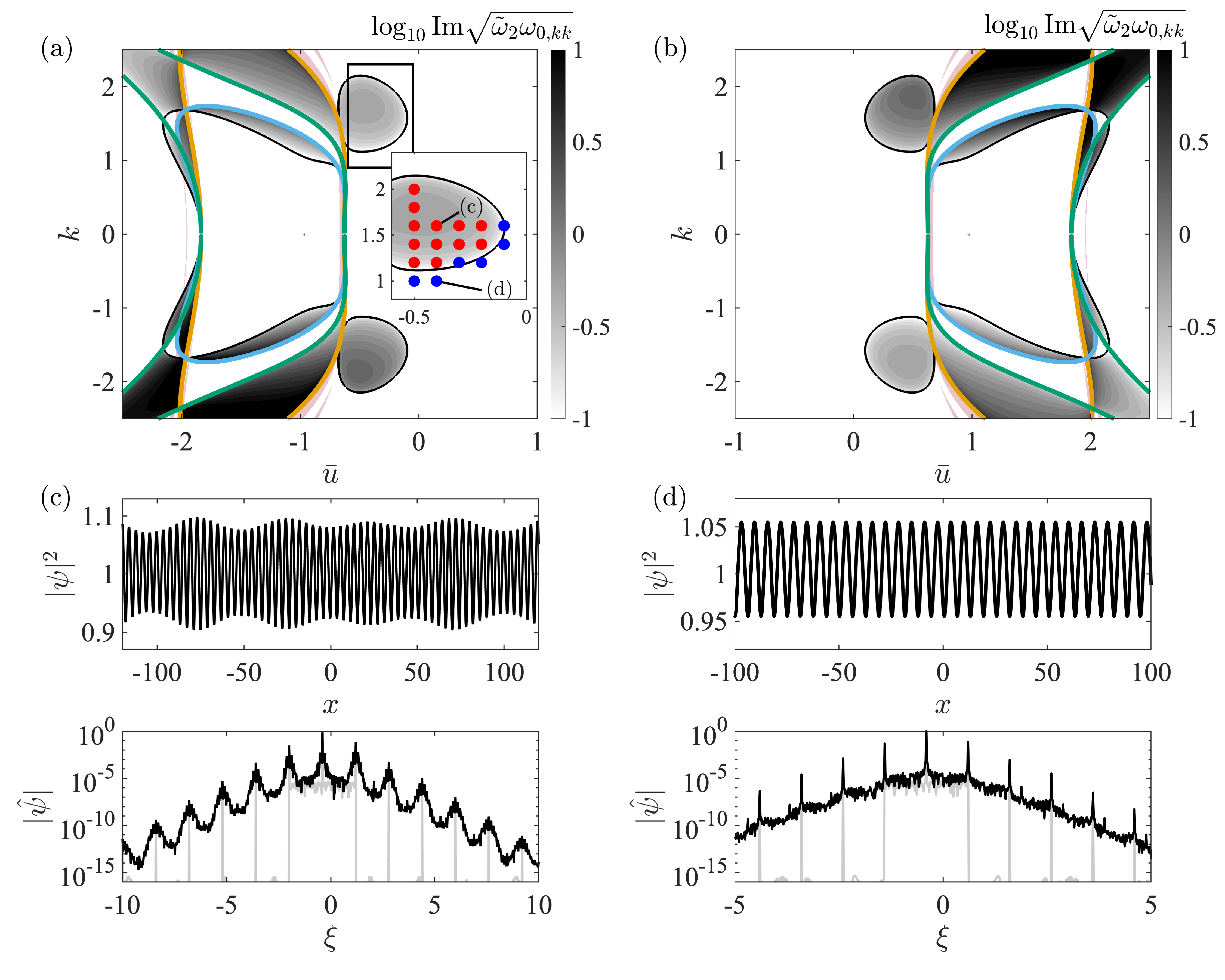}
  \caption{\
    Phase diagrams and numerical simulations of NLS4 two-phase
    solution stability.  Dots in the inset of (a) correspond to
    solutions that are stable (blue) or unstable (red) in direct
    numerical simulations.  See Fig.~\ref{fig:NLS3_examples} and main
    text for details.}
    \label{fig:NLS_4th_examples}
\end{figure}

Figures \ref{fig:NLS_4th_examples}(a) and (b) depict the stability of
fast and slow, respectively, weakly nonlinear two-phase solutions
according to $\SPMI$.  We focus on the gray islands bounded by the
black curves. These regions are particularly notable since their
boundaries correspond to sign changes in the modified nonlinear
frequency shift, $\tilde{\omega}_2$. We thoroughly probe this region
by computing two-phase solutions with $a = 0.075$, while varying
$\ubar$ and $k$ (red and blue dots in the inset of
Fig.~\ref{fig:NLS_4th_examples}(a)).
Perturbed, fast two-phase solutions that numerically exhibit side-band
growth by $t = 2500$ are indicated by red circles.  Those that do not
have blue circles. The island where $\SPMI = -1$ accurately predicts
the instability of two-phase wavetrains, an example of which is shown
in Fig.~\ref{fig:NLS_4th_examples}(c) at $t = 2500$ for the perturbed
two-phase solution with $(\ubar,k) = (-0.4,1.6)$.  The power undergoes
significant long-wavelength modulations from its initial profile.
This is reflected in the spectral intensity by the large side-bands
about the harmonics at wavenumbers $\xi = \ubar + n k$,
$n \in \mathbb{Z}$.
 
Figure \ref{fig:NLS_4th_examples}(d) shows an example of a stable,
fast two-phase solution with $(\ubar,k) = (-0.4,1)$ at $t =
2500$. Neither the power $|\psi|^2$ nor the amplitude spectrum $|\hat \psi|$
indicate any sign of instability.

\subsection{A model of water waves}
\label{sec:ww}

Weakly nonlinear, nearly monochromatic wavetrains in finite-depth
water waves can be modeled by the cubic NLS equation
\cite{ablowitz_solitons_1981}
\begin{align}
  \label{eq:classical_NLS_ww}
  i A_t =  - \frac{1}{2}\frac{d^2 \Omega_{\rm ww,0}}{d\kappa^2}
  A_{xx} + \nu_0 |A|^2 A ~,
\end{align}
where $\Omega_{\rm ww}(\kappa)^2 = \kappa \tanh \kappa$,
$\Omega_{\rm ww,0} = \Omega_{\rm ww}(\kappa_0)$, and
$\nu_0 = \nu_0(\kappa_0)$ is a real constant. The NLS equation
\eqref{eq:classical_NLS_ww} is a narrow-band model, obtained by Taylor
expanding the water waves linear dispersion relation $\Omega_{\rm ww}$
about the carrier wavenumber $\kappa_0$. More accurate models retain
higher-order asymptotic terms \cite{dysthe_note_1979}. We propose the
following, analogous full-dispersion model of finite-depth water waves
\begin{align}
    \label{eq:NLS_ww_model}
  i \psi_t & =  - \left(\cO_{\rm
             ww}(\kappa_0 - i \partial_x) - \Omega_{\rm ww,0} + i
             c_{{\rm g}_0,\rm ww}\partial_x\right) \psi + {\rm sgn }
             \ \nu_0 \ |\psi|^2 \psi~,
\end{align}
where $\Omega_{\rm ww,0}$ and $c_{{\rm g}_0,\rm ww}$ are the
dispersion and group velocities evaluated at $\kappa_0$.

A similar model was considered by Trulsen et
al.~\cite{trulsen_modified_1996} in deep water where the FDNLS
equation \eqref{eq:generalized-NLS} has dispersion
$\Omega(\kappa) = \sqrt{\kappa}$. More recently, Craig, Guyenne and
Sulem \cite{craig_normal_2020} derived a full-dispersion model of
water waves that includes some of the higher-order nonlinear terms
from the Dysthe equation in deep water.

To simplify the presentation, we set $\kappa_0 = 0$, which is
accomplished without loss of generality by redefining the carrier
wavenumber as $\ubar \to \ubar + \kappa_0$. To ensure that the
coefficient of the nonlinear term is positive we require
$\ubar < 1.363$ so that the Benjamin-Feir instability
\cite{benjamin_disintegration_1967} does not occur.  A calculation
reveals that the one-phase solutions are modulationally unstable for
$\ubar < 0$ ($\CPMI = 1$). Therefore, we omit this parameter regime and hence, we consider $0 \le \ubar < 1.363$.  

\begin{figure}[H]
\begin{center}
\includegraphics[width=0.99\textwidth]{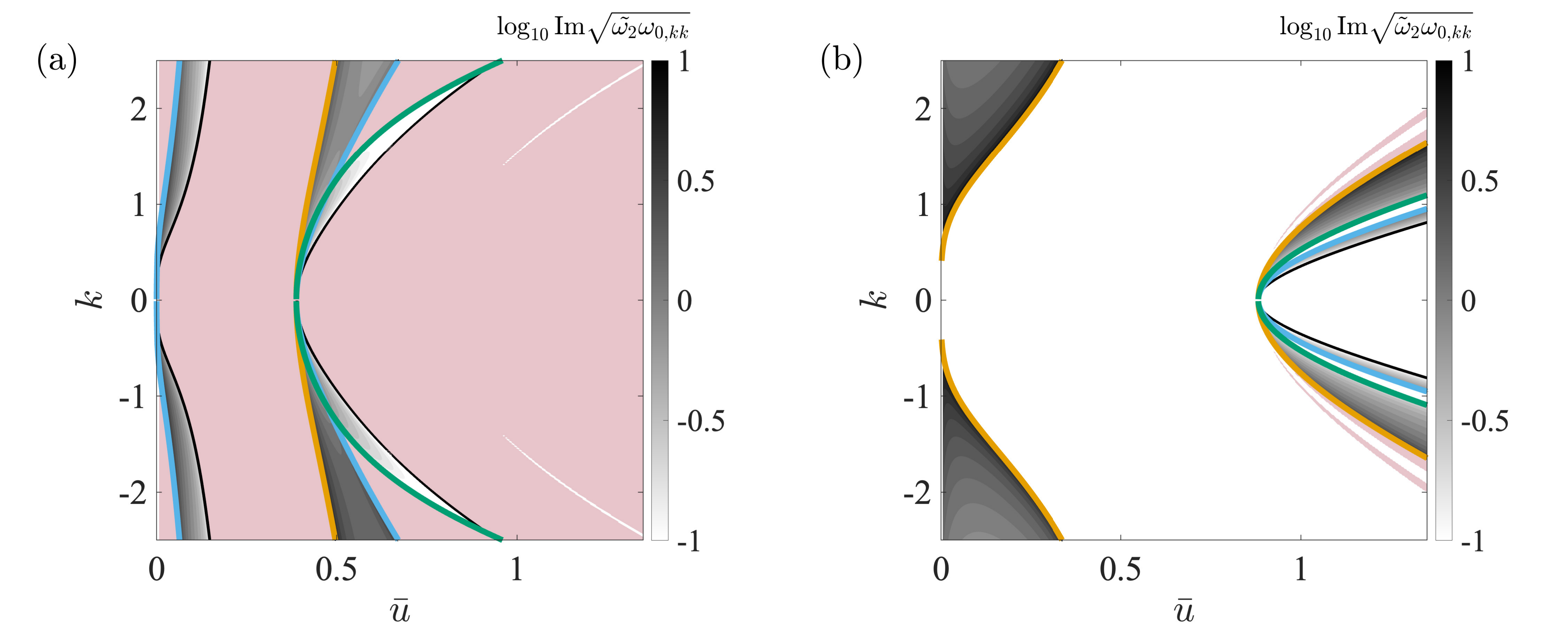}
\end{center}
\caption{
Phase diagrams of two-phase solution stability for
  the water waves model \eqref{eq:NLS_ww_model}.  See
  Fig.~\ref{fig:NLS3_examples} and main text for details.}
\label{fig:stability_waterwaves}
\end{figure}

Figure \ref{fig:stability_waterwaves} presents the stability regions
for weakly nonlinear, fast (a) and slow (b) two-phase wavetrains. The
structure of the stability phase-plane is qualitatively similar to
that of the NLS3 model with third-order dispersion. In the fast
branch, there are high-frequency resonant modes that co-propagate on
the slow dispersion branch, so the resonance mechanism is of the same
nature as in Fig.~\ref{fig:linear_resonances}(a). Evaluation of
$\SPMI$ at the high-frequency resonant modes indicates that an
instability is present, and one can observe this instability in
numerical simulations. In large regions of parameter space, the slow
two-phase waves are stable while narrow (pink) regions exhibit
unstable resonant modes of the type illustrated in
Fig.~\ref{fig:linear_resonances}(b).

\subsection{Discrete NLS}
\label{sec:discNLS}

The defocusing, discrete NLS (DNLS) equation
\begin{equation}
  \label{eq:dNLS}
  i\psi_{n,t} = - \frac{1}{2} \left(\psi_{n-1} - 2
    \psi_n + \psi_{n+1}\right) + |\psi_n|^2 \psi_n,
\end{equation}
has been used, for example, to model the evolution of light in long,
semiconductor waveguide
arrays~\cite{morandotti_dynamics_1999}. Significant attention has been
given to the study of dark solitary wave solutions and their stability
\cite{kivshar_dark_1994,johansson_discreteness-induced_1999}.  To
study two-phase solutions of the DNLS equation, we introduce the
interpolating function $\psi(x,t)$, such that $\psi(n,t) =
\psi_n(t)$. 
The shifting operators in the discrete setting can be understood in a
distributional sense
\begin{align}
    \psi_{n \pm 1} = \psi(n\pm 1,t) = \int_{\mathbb{R}} \delta(x - n \mp 1) \psi(x,t) dx~.
\end{align}
The action of the advance-delay operator in the FDNLS model
\eqref{eq:generalized-NLS} can be represented by its Fourier transform
as
\begin{equation*}
  \begin{split}
    \mathcal{F} \{\cO(-i\partial_x)\psi\} &=
    \mathcal{F}\left\{\psi(n-1,t)- 2 \psi(n,t) + \psi(n+1,t)\right\} \\
    &= 2 \left(\cos(\xi) - 1\right)\hat{\psi}(\xi,t), 
  \end{split}
\end{equation*}
where $\hat{\psi}$ is the Fourier transform of the interpolating
function $\psi$. Thus, we define the continuous DNLS model as
\begin{align}
  \label{eq:continuous-dNLS}
  i\psi_t = \cO(-i\partial_x)\psi + |\psi|^2\psi~
\end{align}
with $\Omega(\xi) = 1 - \cos(\xi)$, whose solution is an interpolant
of the solution of the DNLS equation \eqref{eq:dNLS}.  Here, the
admissible range of the wavenumber is $|\ubar| \leq \pi/2$. The
classical MI index \eqref{eq:classical-MI-index} is
\mbox{$\Delta_{\rm CPMI} = \cos(\ubar)$}. Hence, the admissible
one-phase solutions are linearly stable with respect to periodic
perturbations of any wavenumber\footnote{Since this model is a
  continuum approximation of a discrete system, we may only consider
  perturbations that are band-limited. Therefore, they do not
  oscillate on scales below the lattice spacing. In the normalization
  utilized here, the wavenumber of perturbation, $k$, satisfies
  $|k| < \pi$.} 

\begin{figure}[H]
  \centering
  \includegraphics[width=0.95\textwidth]{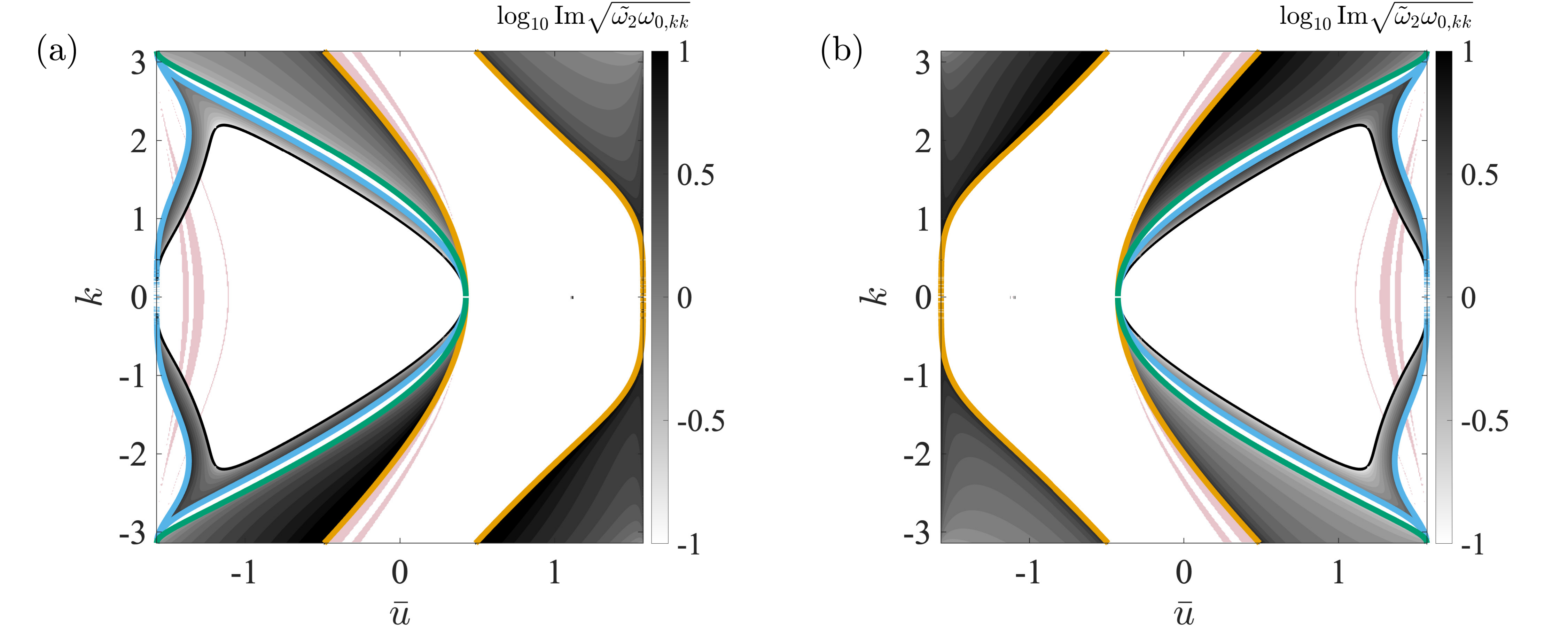}
  \caption{Phase diagrams of two-phase solution stability for the DNLS model
    \eqref{eq:continuous-dNLS}.  See Fig.~\ref{fig:NLS3_examples} and
    main text for details. }
  \label{fig:stability_dnls}
\end{figure}

Figure \ref{fig:stability_dnls} presents the stability diagrams for
fast and slow two-phase solutions of
Eq.~\eqref{eq:continuous-dNLS}. The slow branch resembles that of a
portion of the NLS4 model, Fig.~\ref{fig:NLS_4th_examples}, though a
similar region containing instability islands that we explored in
Sec.~\ref{sec:NLS4} is not permitted in the band-limited region of the
discrete system.

\section{Discussion and Conclusions}
\label{sec:conclusions}
In this manuscript we derived the generalized Whitham modulation
equations \eqref{eq:whitham_mass}--\eqref{eq:cons_theta2} for the full
dispersion nonlinear Schr\"odinger
equation~\eqref{eq:generalized-NLS}.  These equations are obtained by
utilizing a multiscale approach, assuming the existence of two-phase
solutions, and averaging the conservation laws of the FDNLS equation
over the manifold of two-phase solutions.
For weak nonlinearity, the modulation equations' type (hyperbolic /
elliptic) is used to assess the MI of two-phase solutions.  
For the classical cubic NLS equation, the MI of
one- and two-phase wavetrains is directly related.  If a one-phase
solution is modulationally stable (unstable) to long wavelength perturbations, then weakly nonlinear
two-phase solutions with the same carrier wavenumber are also
modulationally stable (unstable). One goal of this study was to
identify scenarios in which the one-phase carrier wave is stable, but unstable when modulated with a finite amplitude second phase. 

We obtain the characteristic velocities   perturbatively 
and identify two distinct indices that determine the reality or
complexity of these velocities.  One index [Eq.~\eqref{eq:classical-MI-index}]  corresponds to the known classical or generalized MI criterion.
The new index [Eq.~\eqref{eq:MI_index_second_phase}] 
determines the MI of two-phase
solutions.  This index depends on the nonlinear
potential $f$ and the linear dispersion function $\Omega(\xi)$ that
define the FDNLS equation as well as the three parameters of the weakly
nonlinear two-phase wavetrain:  mean density $\rhobar$,  carrier-phase wavenumber, $\ubar$, and the second-phase wavenumber, $k$.

The derivation of the modulation equations and the two-phase index are
obtained under general conditions.  The stability predictions are
determined and favorably compared with numerical simulations for the
FDNLS equation with third and fourth-order dispersion. Secondary instabilities were also identified, both theoretically and
numerically, resulting from linear inter- and intra-dispersion branch
resonances. Nonlinearity serves to couple linearly resonant modes with
different wavenumbers and the predicted instability of corresponding
weakly nonlinear two-phase solutions with the same carrier wavenumber
and the resonant wavenumber leads to instability.  This phenomenon is
studied in the FDNLS equation with third-order dispersion, which
exhibits qualitatively similar features with a FDNLS model of water
waves.  Understanding the physical implications of these predictions,
if any, is reserved for future work.

We posit that the FDNLS model provides insight into the nature of high-frequency instabilities.  We
note that high-frequency instabilities were observed in numerical
simulations of water waves ~\cite{deconinck_instability_2011} and have
recently been analyzed via asymptotics and rigorous spectral
methods~\cite{creedon_high-frequency_2022, hur_unstable_2023}.

The generalized Whitham modulation equations can also be used to study
solutions of initial value problems.  In NLS-type equations with
higher-order dispersion, distinct DSW structures emerge that have been
identified with resonances \cite{conforti_resonant_2014}. These DSWs
are particular, in that they may be described in terms of a shock
solution of the Whitham modulation equations, an area of recent
interest in nonlinear dispersive wave equations with higher-order
dispersion
\cite{hoefer_modulation_2019,sprenger_discontinuous_2020,baqer_modulation_2020,baqer_nematic_2021}.

A novel application of the results in this manuscript are in the
modulation of two-phase solutions in discrete systems, which is
accomplished upon casting the advance-delay operators in these systems
as a pseudodifferential operator. A promising direction of research is
to utilize the Whitham modulation theory developed here
to study DSW structures that evolve from step-like initial data in
full-dispersion NLS models, such as the discrete NLS equation.

\section*{Acknowledgements}
The authors would like to thank the Isaac Newton Institute for Mathematical Sciences for support and hospitality during the programme “Dispersive hydrodynamics: mathematics, simulation and experiments, with applications in nonlinear waves” when the work on this paper was completed.

\bibliographystyle{siam}

\end{document}